%% file: main.tex
\documentclass[preprint,11pt]{elsarticle}
\usepackage{geometry}
\journal{Science of Computer Programming}
\usepackage[T1]{fontenc}
\usepackage{amsmath,amssymb,amsthm}
\usepackage{graphicx,xcolor,url}
\usepackage{libertine}
\usepackage[scaled=.92]{inconsolata}
\usepackage{microtype}
\usepackage[libertine]{newtxmath}
\usepackage[hidelinks,bookmarksnumbered]{hyperref}
\providecommand{\Description}[1]{}
\biboptions{numbers,sort&compress}
\usepackage{booktabs}
\usepackage{etoolbox}
\usepackage{longtable}
\usepackage{array}
\usepackage{tabularx}
\usepackage{calc}
\usepackage{tikz}
\usepackage{placeins}
\newtheorem{gtheorem}{Theorem}[section]
\newtheorem{glemma}[gtheorem]{Lemma}

\newtheorem{gdefinition}[gtheorem]{Definition}
\newtheorem{gcorollary}{Corollary}[section]
\usetikzlibrary{arrows.meta,calc,positioning}
\newcommand{\tightlist}{}

\newcounter{none}

\newcounter{gcpalgorithm}
\renewcommand{\thegcpalgorithm}{\arabic{gcpalgorithm}}
\makeatletter
\patchcmd{\@makecaption}{#1: #2}{#1\enspace #2}{}{\PackageError{gcp}{Short caption format changed}{}}
\patchcmd{\@makecaption}{#1: #2}{#1\enspace #2}{}{\PackageError{gcp}{Long caption format changed}{}}
\def\els@aparagraph[#1]#2{\elsparagraph[#1]{#2\@addpunct{.}}}
\def\els@bparagraph#1{\elsparagraph*{#1\@addpunct{.}}}
\newcommand{\gcpresultlabel}[2]{%
  \phantomsection
  \begingroup
  \def\@currentlabel{#2}%
  \label{#1}%
  \endgroup}
\makeatother

\input{supplement-labels.tex}
\begin{document}
\begin{frontmatter}
\title{Generic Constraints Projection: An Exact Criterion for Preserving\\ Type Comparison Reports under Record Width Extension}
\author{Qunhui Zhang\corref{cor}}
\ead{will_zhang@sjtu.edu.cn}
\cortext[cor]{Qunhui Zhang is the corresponding author.}
\affiliation{organization={School of Software, Shanghai Jiao Tong University},addressline={Software Building, 800 Dongchuan Road, Minhang District},city={Shanghai},citysep={},postcode={200240},country={China}}
\begin{abstract}
\input{abstract.tex}
\end{abstract}
\end{frontmatter}
\raggedbottom
\tolerance=1800\emergencystretch=5em
\hypersetup{pdftitle={Generic Constraints Projection: An Exact Criterion for Preserving Type Comparison Reports under Record Width Extension},pdfauthor={Qunhui Zhang}}
\setlength{\tabcolsep}{4.5pt}
\input{body.tex}

\section*{CRediT authorship contribution statement}

\noindent\textbf{Qunhui Zhang:} Conceptualization, Methodology, Software, Validation, Formal analysis, Investigation, Data curation, Writing -- original draft, Visualization.

\section*{Funding}
The author received no funding for this research or for the preparation of this manuscript.

\section*{Declaration of competing interest}
The author declares no competing financial interests or personal relationships that could have appeared to influence the work reported in this paper.

\section*{Declaration of generative AI and AI-assisted technologies in the manuscript preparation process}
The author used OpenAI Codex to translate and revise the manuscript and to check the proofs. The author reviewed and edited the material and takes responsibility for the submitted text and results.
\bibliographystyle{elsarticle-num-names}
\bibliography{refs}

\end{document}

%% file: supplement-labels.tex
\newcommand{\suppnumber}[1]{\csname supplement@#1\endcsname}
\expandafter\def\csname supplement@app:consumer-cost\endcsname{S5.5}
\expandafter\def\csname supplement@app:frame-cost\endcsname{S5.7}
\expandafter\def\csname supplement@app:frame-receiving-cost\endcsname{S5.6}
\expandafter\def\csname supplement@tab:consumer-cost-ranges\endcsname{S18}
\expandafter\def\csname supplement@tab:consumer-stage-ranges\endcsname{S19}
\expandafter\def\csname supplement@tab:consumer-stages\endcsname{S20}
\expandafter\def\csname supplement@tab:frame-allocation\endcsname{S23}
\expandafter\def\csname supplement@tab:frame-cost-all\endcsname{S22}
\expandafter\def\csname supplement@tab:frame-ranges\endcsname{S24}
\expandafter\def\csname supplement@tab:frame-receiving-ranges\endcsname{S21}

%% file: abstract.tex
A release can preserve a program's type and output while changing a comparison cited in an approved report. A publisher needs to know whether the original approval covers the new release's report under the same identifier and which source comparisons yield different reports. Reconstruction answers these questions after projection. We study them in Generic Constraints Projection for a pure record language whose comparison stops at type equality. For a fixed source export over ground types built from immutable records, primitives, top and bottom, we prove an exact criterion under record width extension. Two independently projected valid certificates yield the same report up to session renaming exactly when their read bindings agree and the actual type at each old stopping position is unchanged. With read bindings fixed, we trace these positions to input paths. If each old record admits a fresh field name, the resulting subtree equalities form the unique irredundant conjunction on old input-record paths for whole-report preservation. For completed mixed edits, a separate criterion uses the new valid certificate to check required types, old leaf endpoints and the surplus fields that keep old branches open. Across two consecutive API Extractor release windows, one published member addition changed an interface yet preserved both comparisons in our mapper. Given the old export, input types and fixed read bindings, occurrence-specific paths identify, before reprojection, exactly which source comparisons would yield different reports after proposed field additions.

%% file: body.tex
\input{sections/01-introduction.tex}
\input{sections/02-background.tex}

\input{sections/02b-route-evaluation.tex}
\input{sections/09-related.tex}

\input{sections/10-discussion.tex}

%% file: sections/01-introduction.tex
\section{Introduction}\label{introduction}

\subsection{What a Saved Approval Authorizes}\label{sec:running-example}

A type certificate records a type assignment for a program. Its comparison report records how that assignment satisfies the source constraints. An entry names the source occurrence, the path where comparison stopped and the types compared there. When a release approval covers these entries, carrying it forward requires knowing which of them survive an input change. A program can still type-check and return the same value even though a comparison in the approved report has changed.

Consider an inventory mapper that reads the version range of \texttt{debug} from an Express package manifest. Express 4.17.3 and 4.18.0 both request \texttt{debug} at \texttt{2.6.9}. The latter release also adds \texttt{http-errors}. In our replay the old dependency read selects the complete old dependency row. Comparison stops at \texttt{/field:dependencies} because the actual and required rows are equal. On the new manifest the read again selects that required row. Its wider actual row now compares field by field. The mapper returns the same bytes under the same schema, while the old assertion of whole-row equality no longer holds. On ten other transitions the requested range changed and the complete type report survived (Table~\ref{tab:express-releases}). A review of a dependency value and a review of its type evidence therefore authorize different facts.

Compiler tools already make aspects of successful elaboration observable. A Lean regression test constructs a pair of natural numbers and checks the message that renders the constructed term \citep{lean-show-term}. A change to that message can fail the test even when the example still elaborates at the same type. In release engineering, Rush Stack tracks public API declarations and assigns owners to changes in its review files \citep{rush-report,rush-policy,rush-owners}. Our prototype brings these two concerns together. A release engineer publishes the successful comparisons as an independently inspectable explanation of a mapper. A downstream reader can cite an entry by its position in an approved report. Reattaching the same identifier to a newly expanded comparison would make an old citation denote a different assertion. For this publication contract, the extent of the review follows from what the identifier promises its readers.

We call a stopping position in the canonical comparison a \emph{cut}. Given the old export, capture types and read bindings, our aim is to locate the cuts a proposed insertion would affect if the bindings stay fixed. For a completed revision that may also remove fields or select new read types, we determine which source comparisons retain their old evidence using the new valid certificate.

\subsection{Exact preservation and its consequences}\label{sec:this-paper}\label{the-gcp-framework}

We study report preservation in Outline\(_R\), a pure curried record fragment with parameters, constants, field reads and record construction. Generic Constraints Projection (GCP) exports the obligations of a parsed function, captures its arguments in stages and produces a finite-domain certificate. A checks the supplied residual bounds, B checks the original bounds under the final substitution, and C performs B and checks the submitted report against its canonical reconstruction. Section~\ref{sec:source-residual} gives the conditions under which these checks imply one another.

For a fixed source export, the same session namespace up to renaming, and ground types built from immutable records, primitives, top and bottom, we derive four results.
\begin{enumerate}\tightlist
\item \emph{Identity of the report.} Under width extension, two independently projected valid certificates yield the same report exactly when their read bindings agree and the actual type is unchanged at every old comparison stop (Theorem~\ref{thm:report-frame}). The old leaves recover the required type at each read. Once those choices agree, an old expanding branch cannot close early.
\item \emph{Dependencies on the input.} At fixed read bindings, we trace the old stops back to a set of input paths. If each old record admits a fresh label, the resulting subtree equalities form the unique irredundant conjunction of such tests on old input-record paths over all width extensions (Corollary~\ref{cor:source-footprint}).
\item \emph{Affected source comparisons.} Given the sites of proposed field additions, the paths retained for each occurrence identify exactly which comparisons would change at those bindings, before a new certificate exists. Equation~\eqref{eq:attribution-from-global} characterizes when the outcomes of the smaller whole-report test already determine that set (Corollary~\ref{cor:occurrence-impact}).
\item \emph{Reuse after a completed edit.} For two successful ground comparisons at the same occurrence, the current required type, the endpoints at old leaves and the surplus fields at particular old branches decide whether the old evidence survives. This test uses the new certificate and admits deletions and read reselection (Corollary~\ref{cor:surplus}).
\end{enumerate}

The distinction matters for a single release decision. On a proposed addition, a member diff identifies the interface that changes. The source paths identify which of its comparisons would change. Once a certificate is available, direct reconstruction yields a reference decision for exactly the same report. In API Extractor~7.52.3 the interface acquired a member while both comparisons in our mapper retained their evidence. Conversely, a deletion in Chalk~2.0.0 preserved all three types along the old leaf paths but closed the comparison at their parent. The two cases exercise different parts of the result (Sections~\ref{sec:rush-review} and~\ref{sec:mixed-evaluation}).

\paragraph{Relation to the GCP inference framework.}
The GCP framework supplies the four sources of inference constraints, fresh call projection and the Outline implementation \citep{zhang-gcp-preprint-2026}. Here we attach report entries to source occurrences and characterize when the reports of two independently checked certificates agree. This characterization identifies the field additions that preserve an entire report and, together with a test for mixed edits, the individual comparisons whose evidence survives. The earlier TypeEvalPy study evaluates agreement between inferred types and benchmark labels. Supplementary material~S1 documents that evaluation, its scoring conventions and the remaining difference from the preprint.

\subsection{Technical Scope}\label{claim-scope}

Table~\ref{tab:claim-correspondence} identifies the assumptions behind the source route and its supporting components. The source route goes from a parsed function through export $E$, residual template $T_\rho(E)$ and a finite certificate to B/C checking and delivery.

\begin{table}[t]
\centering\small
\caption{Guarantees of the GCP components}\label{tab:claim-correspondence}
\begin{tabularx}{\linewidth}{@{}p{.19\linewidth}XX@{}}
\toprule
Component & Guarantee & Scope\\
\midrule
Outline\(_R\) source route & Original bounds give value coverage. Capture preserves the report for a fixed assignment. The width-extension criterion characterizes report equality after independent projection. A local test covers successful ground comparisons after mixed edits. & Fixed source export. For report preservation, capture and comparison types lie in the ground record fragment. Delivery also checks the exact output shape.\\
Finite template kernel & Certificate validity, termination, and relative completeness without a budget cutoff. & Finite, explicitly supplied candidate domains (Section~\ref{sec:calculus}).\\
Recursive carrier & A bounded adapter can supply candidate types to the finite kernel. & The article's source theorem uses the direct export of Outline\(_R\). The broader carrier and its benchmark are documented in S1.\\
\bottomrule
\end{tabularx}
\end{table}

The delivery interface supports three units of approval. A named equality binds its endpoints. A canonical leaf also binds its occurrence and decomposition path. A complete report binds the ordered entries at every occurrence. Section~\ref{sec:approval-contract-choice} defines these contracts and the output checks used at delivery.

\paragraph{Reading guide.}\label{paper-organization}
Section~\ref{sec:contracts} develops the four preservation results and the delivery contract. Section~\ref{sec:route-evaluation} follows the report from an input dependency to the affected source occurrences and tests reuse on completed releases. It then measures the receiving workflow. Supplementary material S1 gives the detailed checker controls and timing distributions.

%% file: sections/02-background.tex
\section{Source Contracts, Report Preservation and Checked Execution}\label{background-the-outline-carrier-language-and-its-type-space}\label{sec:contracts}

The source export gives each comparison an identity that survives staged argument capture (Section~\ref{sec:source-residual}). This lets us ask whether the comparison itself survives an input edit. Section~\ref{sec:report-frame} proves the preservation criterion and traces its tests back to the program inputs. Sections~\ref{sec:calculus}--\ref{sec:bridge} describe how to produce a valid certificate. Sections~\ref{sec:checked-execution}--\ref{sec:consumer-contract} specify how a receiver uses it to check the evidence and the output it is asked to deliver.

\input{sections/support/source-residual.tex}
\input{sections/support/report-frame.tex}
\input{sections/support/finite-kernel.tex}
\input{sections/support/checked-execution.tex}
\input{sections/support/consumer-contract.tex}

%% file: sections/support/source-residual.tex
\subsection{Source Export and Residual Origin Preservation}\label{sec:source-residual}

Consider \texttt{r -> s -> \{a=r.k,b=r.k\}}. Capturing \texttt{r} leaves a function of \texttt{s}; both reads still belong to the original body. A residual producer might omit a necessary inequality, or merge the two occurrences because their ground inequalities coincide. These defects require different checks: rechecking every original inequality establishes semantic admissibility, while checking a submitted occurrence trace additionally establishes which source reads that final evidence represents. Theorem~\ref{thm:residual-origins} establishes these two guarantees for the source export.

\paragraph{Source fragment}
Outline\(_R\) consists of distinct unary parameters and a pure body:
\[
 f=x_1\to\cdots\to x_n\to e,
 \qquad e ::= x\mid c\mid e.k\mid\{k_i=e_i\}_{i\in I}.
\]
Constants are signed 32-bit integers, strings and booleans; records are immutable and may be empty, field names are distinct, and variables name prefix parameters. The displayed grammar defines the complete body syntax. Actual values are ground constants or finite immutable records, supplied in one or several stages. The mathematical empty record \(\{\}\) is written \texttt{\{,\}} in executable Outline, where \texttt{\{\}} denotes an empty block. Integer literals include a leading minus: typing and the frozen expression snapshot normalize literal negation and enforce the range $-2^{31}$ through $2^{31}-1$. The value API accepts Java integers and immutable records. Export supports arbitrary finite record depth; the persisted-claim codec's resource limit is specified in Section~\ref{sec:saved-claim}. Caller-supplied finite ground candidate domains determine the search space of the kernel in Section~\ref{sec:calculus}.

\paragraph{Source export}
Give parameter \(x_i\) variable \(p_i\). Before carrier inference, traverse the parsed body using \(\Gamma\vdash e\rightsquigarrow(\tau,O)\), where \(O\) is an ordered list of origin-indexed inequalities. Variables return their environment type and constants their base type, with no obligations. A record exports each child and combines the child types into a row, concatenating obligations in field-name order; the empty record returns \((\{\},[])\). For a field read at AST path \(o\), first export its host as \((\tau,O)\), allocate fresh \(r_o\), and return
\[
 \Gamma\vdash e.k\rightsquigarrow
 \bigl(Ref(r_o),\ O\mathbin{+\!+}
 [(o,\mathsf{structure},\tau\preceq_K\{k:Ref(r_o)\})]\bigr).
\]
The ground relation \(\preceq_K\) includes record width/depth and primitive compatibility; its complete executable definition appears in Section~\ref{sec:subtype}. Each root stores its expression and available source offsets. AST paths distinguish occurrences even when offsets or substituted endpoints coincide. All allocated variables receive the supplied finite domain. The immutable export is \(E=(\bar p,D,O,\tau_r,e)\). Each selected field supertype defines a coverage guarantee for the result.

\paragraph{Residual specialization}
A session \((j,E,\rho)\) contains a fresh namespace \(j\) and a ground substitution for a captured prefix. Binding immutable value \(v_i\) sets \(\rho(p_i)=type(v_i)\), after its domain-membership check. The residual retains every original root, partially substitutes its endpoints and return, and removes captured variables from the search domains. It retains ground true roots as well as unresolved ones. An incomplete prefix yields a residual closure; evaluation of the frozen body follows a successful projection after all actuals arrive. The declared producer constructs exactly this root-retaining residual.

\paragraph{Concrete kernel interface}
For a completed capture, define
\begin{align*}
 V_\rho&=\operatorname{dom}(D)\setminus\operatorname{dom}(\rho),
 &\mathcal D_\rho&=D|_{V_\rho},\\
 C_\rho&=[(\rho L,\rho U,\mathsf{structure},\lambda(o))]_{(o,L,U)\in O},\\
 T_\rho(E)&=(V_\rho,\mathcal D_\rho,C_\rho,[],\mathsf{Unit},\rho\tau_r).
\end{align*}
Candidate order is unchanged. Each \(\lambda(o)\) retains the AST path and available offsets; the residual root wrapper retains \(o\) and its expression. Selectors are empty. All parameter actuals have already been checked and captured, so the remaining kernel call is \(\operatorname{project}(T_\rho(E),\mathsf{Unit})\): its actual/formal check is trivial. A successful certificate \((\eta,R)\) has \(\operatorname{dom}(\eta)=V_\rho\), giving \(\theta=\rho\uplus\eta\) and \(R=\eta(\rho\tau_r)\). Java's \texttt{Session.template}/\texttt{project}/\texttt{extend} implement this interface. The \texttt{template} method consumes the supplied residual bounds \(S\); the declared root-retaining producer supplies \(C_\rho\).

\paragraph{Three checker contracts}
All three checkers require \texttt{Success}, a complete captured prefix, exact variable/domain membership, capture types equal to the actual immutable value types, and reported return \(\theta\tau_r\). Holding these conditions fixed, A checks only the supplied residual inequalities; B checks \(\theta L\preceq_K\theta U\) for \emph{every original} root \((o,L,U)\in O\); C performs B and compares a supplied trace with the deterministic reconstruction from \(E,\theta,j\). Thus C implies B, and B already suffices for source evaluation safety. The original export is the trusted reference. C reconstructs the report from the original roots. If a supplied residual \(S\) drops a root but \(\theta\) still satisfies the original bounds and the submitted trace remains complete, C accepts. To verify the stored residual itself, a receiver must additionally check \(S=\rho O\) or compare its decomposition with the submitted trace. The implementations are \texttt{checkResidual}, \texttt{checkOriginal} and \texttt{check} respectively.

\paragraph{Trace semantics}
Write \(\mathcal D(j,o,p,L,U)\) for the ordered decomposition leaves of a ground comparison. Each leaf records the session/root pair, decomposition path, both endpoints and truth value. The implementation encodes the pair as \(\operatorname{esc}(j)\mathbin{/}o\), where \(\operatorname{esc}\) replaces \texttt{\textasciitilde} by \texttt{\textasciitilde0} and \texttt{/} by \texttt{\textasciitilde1}. The first literal slash uniquely separates the namespace from the source root, including for an empty namespace. Thus \((\texttt{review},\texttt{body/body})\) and \((\texttt{review/body},\texttt{body})\) remain distinct. Trace production, original-root checking and explanation lookup use this same encoding. Paths are sequences of constructor edges. Their string encoding replaces \texttt{\textasciitilde} by \texttt{\textasciitilde0}, then \texttt{/} by \texttt{\textasciitilde1}, inside each field label before appending the edge. This injective encoding distinguishes a single label \texttt{a/field:b} from successive fields \texttt{a} and \texttt{b}; ordinary identifiers are unchanged. Equal pairs and top/bottom shortcuts emit one successful leaf. Otherwise, record comparison recurses through each required right-hand field, appending \texttt{/field:k}; a missing field emits a false leaf. Crucially, the empty required row has an explicit successful witness:
\[
 \mathcal D(j,o,p,\{k_i:L_i\}_{i\in I},\{\})
   =[(j,o,p,\{k_i:L_i\}_{i\in I},\{\},\mathsf{true})].
\]
This also covers distinct endpoints, such as \(\{z:Int\}\preceq_K\{\}\). Arrays recurse covariantly. Arrows reverse the domain comparison, recording \texttt{/domain-}, and preserve the result direction with \texttt{/result+}. Other ground predicates, including Option coverage and invariant dictionaries, remain atomic checked leaves. C requires exact ordered-list equality and valid leaves; root identity and multiplicity are part of equality. If \(O=[]\), the correct complete trace is empty.

\begin{gtheorem}[Residual obligation and origin preservation]\label{thm:residual-origins}
Let \(E\) export Outline\(_R\), let \(\rho\) be a well-formed captured prefix, and let \(\eta\) assign exactly its remaining variables within their domains. Then:
\begin{enumerate}
\tightlist
\item \(\eta\models\rho O\) iff \(\rho\uplus\eta\models O\), and
\(\eta(\rho\tau_r)=(\rho\uplus\eta)\tau_r\).
\item Decomposing \(\eta(\rho O)\) yields exactly the origin-indexed leaves obtained from \((\rho\uplus\eta)O\). Every original root has at least one leaf, and every leaf names one original root. The declared root-preserving capture and fresh-copy operations preserve these properties. For a rootless export the trace is empty.
\item If the final B check succeeds and immutable actuals have the captured types, evaluation terminates without a missing-field error and returns a value covered by \(\theta\tau_r\). C additionally certifies the submitted trace's agreement with the original occurrence-indexed decomposition, including multiplicity, paths and endpoints.
\end{enumerate}
\end{gtheorem}
\subsubsection{Proof of residual origin preservation and value coverage}\label{proof:residual-origins}

We prove Theorem~\ref{thm:residual-origins} for the explicitly exported grammar of Section~\ref{sec:source-residual}.

\paragraph{Substitution composition}
The captured map \(\rho\) has ground ranges and disjoint domain from \(\eta\). Induction on a symbolic type gives \(\eta(\rho\tau)=(\rho\uplus\eta)\tau\). A captured variable uses the same ground range on either side, a remaining variable uses \(\eta\), and every constructor rebuilds the identical children. Applying this equality to both endpoints of each indexed root proves part (1), including the return equality. Removing captured variables from the search domains is exact because their fixed values have already passed their original domain checks. No requirement is discharged merely by removing its variable from the domain map.

\paragraph{Origins and decomposition}
Substitution changes endpoints, not the root list, order or identifiers. By the composition equality, both decompositions start with syntactically identical ground endpoints and root IDs. Induction on the deterministic decomposition tree therefore gives identical leaves and signed paths. Splitting at the first literal slash and decoding the namespace recovers the unique session/root pair. Decoding escaped field labels likewise recovers the unique edge sequence. Equal encodings therefore preserve both occurrence identity and constructor paths. A nonempty required record row has at least one recursive child; each child preserves its root and appends its field. An empty required row emits the explicit successful leaf with the original root, path and both endpoints, even when the left row is nonempty. Arrays have one child; arrows have two and record the domain reversal; atomic, equal and sentinel cases emit one leaf. Missing fields emit a false leaf with the same root. Thus no root becomes leafless, and no decomposition step invents a root. Each finite tree is rooted in exactly one original occurrence. If there are no original roots, there are no trees and the whole trace is correctly empty; the per-root assertion is then vacuous. This proves part (2). Another capture composes a disjoint ground binding with \(\rho\), so the same argument applies at every prefix. Fresh copying changes only the session namespace, preserving original root IDs and every endpoint. It cannot modify another session's captured map.

\paragraph{Ground trace validity}
Every decomposition step preserves the truth of its input inequality: record subtyping is the conjunction of required-field comparisons, whose empty conjunction is represented by a true leaf rather than an absent root; arrays are covariant, and arrows reverse the domain comparison. Equal pairs and top/bottom shortcuts hold by the ground order. Other predicates are checked atomically. A missing field is false. Induction gives validity of a root if and only if all its leaves are true. B checks this same truth condition directly on every original ground inequality, without lineage. C adds exact equality to the reconstructed occurrence list. Therefore both reject a substitution violating an omitted original bound, while only C rejects an otherwise valid certificate with omitted, duplicated or reassigned leaves. Both checks trust the original export. Part (2) applies to the constructed residual \(\rho O\). C's comparison is reconstructed from \(O\) and therefore remains true if \(S\) loses a root but the certificate retains the full original trace and an original-valid substitution. No step of the value-coverage proof uses \(S\).

\paragraph{Source evaluation}
Strengthen part (3) to each export judgment: if \(\theta\) satisfies its roots and the value environment covers the parameter types, the expression evaluates to a value of a subtype of its exported type under \(\theta\). Variables use the environment; literals use their base types. Record construction applies the induction hypothesis to each field and then width/depth subtyping; for the empty constructor it produces the empty record with its exact type. Identity and constant bodies generate no roots and use these same variable/literal cases without a nonempty-trace premise. For \(e.k\), induction gives a value \(v\) with \(type(v)\preceq\theta\tau\). The generated original root supplies \(\theta\tau\preceq\{k:\theta Ref(r_o)\}\). Transitivity gives a record value containing \(k\), whose field type is below \(\theta Ref(r_o)\); lookup is defined and has the required result type. This remains true for nested reads. Evaluation follows the finite expression tree, so it terminates. Each prefix application extends the immutable environment. Body evaluation starts after capture is complete. These cases exhaust the grammar and prove the semantic clause of part (3) using B alone. C implies B and adds exact occurrence-list equality, proving the remaining clause. In particular, equality of two substituted inequalities cannot justify merging their distinct original root IDs.

\paragraph{The extra role of lineage}
In \texttt{r -> s -> r.k}, capturing \texttt{\{other=1\}} and deleting the root may let A accept; B and C reject. The opening two-read example instead has distinct \texttt{body/a} and \texttt{body/b} roots with equal ground constraints. Coalescing their leaves passes B but fails C and a source-ID multiset check. To isolate payload fidelity, project \texttt{r -> r.k} at \texttt{\{k=7,extra=true\}} selecting \(Int\). Its leaf records \texttt{/field:k} and \((Int,Int)\). Changing only the path to \texttt{/field:extra}, or only the endpoints to \((Bool,Bool)\), preserves a true comparison and the root count. An ID/count check accepts; C rejects both.

%% file: sections/support/report-frame.tex
\subsection{Report Preservation under Width Extension}\label{sec:report-frame}

Unused fields can change an explanation even when the selected read types stay fixed. For \texttt{r -> r.k} with field type \(Int\), the actual type \(\{k:Int\}\) equals the required type, so decomposition emits a whole-record leaf at the root. Extending the actual to \(\{k:Int,note:Bool\}\) instead emits \(Int\preceq_K Int\) at \texttt{/field:k}. Both certificates pass B, but the reports differ. A reuse condition must account for this stopping policy as well as the fields demanded by the program. Figure~\ref{fig:unread-field-report} shows that both executions return \(7\). Adding \texttt{note=true} changes the stopping path from the root to \(p=k\) (\texttt{/field:k}) at the same successful source comparison.

\input{figures/fig-09-unread-field-report.tex}

We characterize this dependence for the read-only ground fragment
\[
 t ::= b\mid\bot\mid\top\mid\{k_i:t_i\}_{i\in I},
\]
where \(b\) is a primitive kernel type. The theorem constrains instantiated comparison types in addition to the source grammar. A source read can select an \(\mathsf{Option}\) candidate in the broader kernel of Section~\ref{sec:calculus}. Write \(a\unlhd a'\) for \emph{width extension}: a non-record type is unchanged, while a record retains every old field, recursively width-extends its type, and may add fields. Thus this relation permits additions inside existing records, but never removes a field or changes a primitive type.

For a successful old comparison \(a\preceq_K u\), define its \emph{cut footprint} \(K(a,u)\), a set of paths of field labels:
\[
 K(a,u)=
 \begin{cases}
  \displaystyle\bigcup_{k\in fields(u)} k\cdot K(a.k,u.k),
       &\begin{array}{l}a,u\text{ are records},\ u\ne\{\},\ a\ne u;\end{array}\\[2pt]
  \{\epsilon\},&\text{otherwise}.
 \end{cases}
\]
This definition uses the old types only. At a proper record comparison it follows the required fields; at an equality, empty-row, primitive or top/bottom cut it protects the entire remaining left subtree. Write \(a|_p\) for that subtree. New fields outside these protected subtrees may vary freely. In particular, \(K(\{k:Int,extra:Bool\},\{k:Int\})=\{k\}\), whereas \(K(\{k:Int\},\{k:Int\})=\{\epsilon\}\).

\begin{gtheorem}[Exact report preservation after projection]\label{thm:report-frame}
Let \(a,u,a'\) belong to the displayed ground fragment, with \(a\preceq_K u\) and \(a\unlhd a'\). Then \(a'\preceq_K u\), and, for fixed root and session,
\[
 \mathcal D(j,o,\epsilon,a,u)=\mathcal D(j,o,\epsilon,a',u)
 \quad\Longleftrightarrow\quad
 \forall p\in K(a,u).\ a|_p=a'|_p.
\]
For arbitrary successful comparisons in this fragment, equal reports for the same root and session imply equal right endpoints.

Now consider two completed B-valid inputs for the same Outline\(_R\) export with ground types in this fragment. Let each captured parameter type be width-extended. After session renaming, the complete report is preserved exactly when every read binding is unchanged and the displayed footprint condition holds at every original root. The statement applies to two independently projected certificates. If the new input satisfies the shared status, capture/domain and return conditions and keeps the old read bindings, B validity follows from the old input. For a rootless export both tests are vacuous.
\end{gtheorem}

\begin{proof}
Width extension in the stated ground fragment is reflexive and transitive and preserves existing paths. We first establish two properties.
\begin{enumerate}
\tightlist
\item If \(a\preceq_K u\) and \(a\unlhd a'\), then \(a'\preceq_K u\).
\item If \(a\preceq_K u\) and \(a\unlhd u\), then \(a=u\).
\end{enumerate}
For (1), a non-record left type is unchanged; a top right type accepts every extension. Otherwise successful record comparison requires all right-hand fields. Extension retains those fields and the induction hypothesis establishes their comparisons. For (2), a non-record width relation is equality. For records, width extension gives \(fields(a)\subseteq fields(u)\), while successful subtyping gives the reverse inclusion. The field sets coincide and the induction hypothesis equates every field type. Bottom, top and primitive comparisons create no additional width-extension case.

It follows that a proper old record comparison stays proper: if \(a\ne u\), \(a\preceq_K u\) and \(a\unlhd a'\), equality \(a'=u\) would contradict (2). The current decomposition therefore visits exactly the same required children until it reaches an old cut. This is the step that excludes an unaccounted change of dispatch before the protected frontier.

We now prove the footprint equivalence by induction on the old comparison. At a proper, nonempty record comparison, both decompositions concatenate the same ordered field groups. Each group has a distinct field-path prefix, and the induction hypothesis makes its equality equivalent to the footprint test on that child. Prefixing and taking the union gives \(K(a,u)\). At any cut, an unchanged \(a\) trivially gives the same leaf. Conversely, if the old leaf and the new report agree, either the new comparison is also a leaf at the same path, in which case endpoint equality gives \(a'=a\), or it expands into required fields. The latter produces only strict descendant paths and cannot equal the old leaf. Thus equality is equivalent to preservation of the whole left subtree protected by \(\epsilon\). An expanded nonempty record comparison has at least one child, and each successful child emits at least one leaf. An empty required row instead emits an explicit witness leaf. These cases complete the equivalence.

Recover the upper endpoint recursively from the paths and endpoints for one root. At a node with a leaf, its stored right endpoint is the subtree at that path. Otherwise a successful comparison in this fragment expands a nonempty record; group the leaves by their next field edge, remove that edge and recover the right subtree for each group. Each required field has a nonempty group because every successful branch emits at least one leaf. These are precisely the fields of the right record. Injective path encoding makes this reconstruction unique. Equal reports therefore give equal upper endpoints even when both certificates may change.

Each source read has one root with upper type \(\{k:\theta(r_o)\}\); equality of reports implies equality of those bindings. With the bindings fixed, all right endpoints stay fixed. Exported lower types contain only parameter/read references, constants and record constructors. Structural induction shows that width-extending parameters then width-extends every lower endpoint. Property (1) preserves all inequalities, and the shared checks give B validity. Apply the ground equivalence at each root. Distinct occurrence IDs separate their ordered leaf lists even when some endpoints coincide. Session renaming changes only the shared prefix, and Theorem~\ref{thm:residual-origins} removes dependence on the partition into capture stages. The rootless report is empty.
\end{proof}

Figure~\ref{fig:report-preservation-test} gives the resulting decision procedure. A changed read binding changes a recoverable right endpoint. When the bindings agree, the receiver compares the protected left subtrees. At an equality cut or an empty required row this includes the entire remaining record, even its unread fields. The test covers every original root after session renaming. Delivery also checks the bound submission, output shape and encoding (Theorem~\ref{thm:capture-migration}).

\input{figures/fig-10-report-preservation-test.tex}

\begin{gcorollary}[Record-valued cuts suffice]\label{cor:record-cuts}
Under the fixed-right-endpoint hypotheses of Theorem~\ref{thm:report-frame}, put
\(K_R(a,u)=\{p\in K(a,u)\mid a|_p\text{ is a record}\}\).
The two reports are equal exactly when \(a|_p=a'|_p\) for every \(p\in K_R(a,u)\).
\end{gcorollary}
\begin{proof}
Width extension preserves old paths and restricts to a width extension at each such path. At a non-record left endpoint, width extension is equality, so every omitted test holds by hypothesis. Apply Theorem~\ref{thm:report-frame}.
\end{proof}

The reduction can remove every cut. For \(\{k:Int,x:Bool\}\preceq_K\{k:Number\}\), \(K=\{k\}\) and \(K_R=\varnothing\). The remaining record cuts have the following exact interpretation.

\begin{gcorollary}[Locations and irredundancy of record-cut tests]\label{cor:cut-locations}
Fix a successful ground comparison \(a\preceq_K u\) in Theorem~\ref{thm:report-frame}'s fragment. Let \(q\) be a path to a record in \(a\), and obtain \(a'\) by adding one fresh field whose type is in the same fragment at \(q\), leaving all other old fields unchanged. The report changes exactly when some \(p\in K_R(a,u)\) is a prefix of \(q\), including equality of paths.

If a fresh field name is available at every record cut, each test in \(K_R(a,u)\) is necessary: for every \(p\in K_R(a,u)\), a successful width extension can preserve the left subtrees at all other record cuts while changing the report.
\end{gcorollary}
\begin{proof}
The cut paths are prefix-free: decomposition stops at each cut and never visits its descendants. Adding a fresh field at \(q\) changes an old cut subtree precisely when that cut is at or above \(q\); all old paths remain defined. Corollary~\ref{cor:record-cuts} proves the first claim. For the second, add a fresh Boolean field at the chosen cut \(p\). No other cut is an ancestor or descendant of \(p\), so its subtree is unchanged. Width extension preserves success, while the changed subtree at \(p\) changes the report by Theorem~\ref{thm:report-frame}.
\end{proof}

The prefix test classifies field additions relative to a ground comparison. We next combine these local tests through the source export to obtain a shared condition on the captured inputs.

\input{sections/support/source-footprint.tex}

\paragraph{Field removal and read reselection.}
Let \(a=\{k:Int,x:Bool\}\) and \(u=\{k:Int\}\), so the old cut is \(k\). Deleting \(x\) leaves the protected subtree \(Int\) unchanged but replaces the field leaf by root equality. The cut test alone passes although the report changes; this edit violates the width-extension premise. Separately, keep \(a\) fixed and select \(Int\) or \(Number\) as the read binding of \texttt{r -> r.k}. Both witnesses are valid, and both reports stop at \(k\) with left endpoint \(Int\), but their right endpoints differ. This unchanged capture satisfies width extension. Thus equality of read bindings is required in addition to the protected-left-subtree tests.

\paragraph{Whole-record equality and field equality}
Take \(u=\{k:Int\}\), \(a=u\) and \(a'=\{k:Int,note:Bool\}\). The required-field observation is \(Int\preceq_K Int\) in both cases. Any authorization decision receiving that observation and the read identity gets identical inputs. Yet a saved assertion of \(a=u\) holds for the first pair and fails for the second. Storing the equality assertion explicitly preserves the distinction. In our canonical report the root leaf and its endpoints retain it, and the footprint theorem characterizes the edits that preserve the report. A policy that always expands nonempty records instead observes required-field assertions. These policies preserve different evidence, made explicit by the approval contracts in Section~\ref{sec:approval-contract-choice}.

\paragraph{Factoring a trace-replay check.}\label{sec:replay-factorization}
Replay of the old comparison tree tests its internal branches and its terminal endpoints. Fix \(u\), and let \(P(a,u)\) contain all strict prefixes of paths in \(K(a,u)\). These are exactly the old expanding nodes. Define \(G_P(a',u)\) to require, at each \(q\in P(a,u)\), two unequal records, a nonempty required row, and all required fields on the left. Undefined paths fail the guard. Let \(Q_K(a,a')\) require defined, equal left endpoints at every old cut. For successful comparisons in the displayed ground fragment, replay has the following factorization:
\[
 \mathcal D(j,o,\epsilon,a,u)=\mathcal D(j,o,\epsilon,a',u)
 \quad\Longleftrightarrow\quad G_P(a',u)\ \land\ Q_K(a,a').
\]
Indeed, the guards recreate the old internal nodes, and endpoint equality recreates their terminal leaves. Conversely, equal reports have the same leaf paths and endpoints. Every old internal node is a strict prefix of a leaf path and must still expand. Each successful child emits a leaf, including an explicit witness for an empty required row, so no branch is hidden from this argument.

The complete prefix test can be reduced even when an edit removes fields. Let \(S(a,u)\) comprise the old expanding nodes \(q\) at which every required child of \(a|_q\) equals the corresponding child of \(u|_q\). Such a node expanded because it had at least one additional field. Its required children were all equality leaves. Write \(Q_K(a,a')\) as above, including cuts with primitive endpoints. A test at an undefined path fails.

\begin{gcorollary}[Surplus fields and reuse after mixed edits]\label{cor:surplus}
In the ground record fragment of Section~\ref{sec:report-frame}, let \((a,u)\) and \((a',u')\) be two successful comparisons at the same occurrence in a common session. Their canonical reports agree exactly when
\[
 u'=u,\qquad Q_K(a,a'),\qquad
 \forall q\in S(a,u).\quad
 \bigl|\mathrm{fields}(a'|_q)\bigr|>\bigl|\mathrm{fields}(u|_q)\bigr|.
\]
The last test requires a record at \(q\). For unrestricted successful changes of the left endpoint at fixed \(u\), each listed surplus-field test has a deletion witness that passes all the other tests. Under width extension all listed surplus-field tests hold automatically.
\end{gcorollary}
\begin{proof}
Equal reports recover the same right endpoint by Theorem~\ref{thm:report-frame}. They visit the same cuts with equal left endpoints. At a node in \(S(a,u)\), each required child still equals its required type. To expand there, the new record must therefore have an additional field.

Conversely, fix the common right endpoint. At an old cut the endpoint test reproduces the leaf. Work upwards through the old expanding nodes. The defined paths to the cuts ensure that the new left record contains every required field. If a required child formerly expanded, induction gives two unequal records there, which makes the parent endpoints unequal as well. If a child was an unequal leaf, equality of its old and new left endpoints preserves that inequality. In the remaining case every child was an equality leaf. The parent belongs to \(S(a,u)\), where the field-count test keeps the records unequal. In each case comparison therefore expands and reproduces the child reports in field order.

For a witness at \(q\in S(a,u)\), delete all surplus fields at that node alone. The comparison stays valid. Every required child and every old cut endpoint stays fixed. The nodes of \(S(a,u)\) are pairwise incomparable since their required children are leaves. All other surplus tests thus pass, while the report loses the old expansion at \(q\) or an ancestor. For a width extension, each old surplus field is retained. This proves the last claim.
\end{proof}
The test uses the number of fields, so changing the name or type of a surplus field can preserve the report. In \(\{k:Int,x:Bool\}\preceq_K\{k:Int\}\), replacing \(x:Bool\) by \(y:String\) preserves the field leaf at \(k\). Deleting \(x\) alone collapses it to root equality. This witness concerns instantiated endpoints; an application's capture domains may restrict which witnesses its source program can realize.

For two valid certificates of the same export, the receiver can instantiate the endpoints for each occurrence and apply Corollary~\ref{cor:surplus} separately, including after read reselection. A changed right endpoint settles its own comparison immediately. Other occurrences using that binding in a lower expression are checked on their own endpoints. For width extensions with fixed reads, Theorem~\ref{thm:report-frame} discharges all expansion tests and Corollary~\ref{cor:source-footprint} moves the remaining tests back to input paths. The resulting source paths predict the effect of additions from their sites before the next certificate exists, conditional on the read bindings being retained.

\subsubsection{Report preservation, trace reuse and slicing}\label{sec:preservation-related}
Table~\ref{tab:evidence-after-edits} groups methods by the evidence they retain, the edits they admit and the resulting guarantee. Reconstructing the current report and comparing it with the saved report decides identity after projection. For field additions, the cut criterion expresses the same identity in terms of the old stopping positions.
From-scratch consistency in Nominal Adapton equates an updated result with a fresh evaluation on the new input \citep{hammer-names-2015}. If the computation produces a report, the guarantee identifies the updated output with the current canonical report. Fidelity in \emph{Provenance Traces} relates successful trace adaptation to evaluation on the changed input \citep[Section~5]{cheney-provenance-traces-2008}. Our observation is the ordered sequence of source comparisons across two inputs. In Section~\ref{sec:running-example}, the returned value is constant while comparison at the dependency read changes from whole-record equality to field decomposition.

\paragraph{Slicing the report as an output.}
Take the complete report as the output pattern. In the query calculus of \citet[Theorem~4.4]{cheney-query-slicing-2014}, a changed input and trace that match their respective slices reproduce the selected output. Their query-slicing result also gives a condition before recomputation. An input agreeing on the input slice and a query extending the query slice produce an output agreeing on the pattern whenever evaluation succeeds \citep[Corollary~5.2]{cheney-query-slicing-2014}. For our fixed export and read bindings, agreement on the input paths of Corollary~\ref{cor:source-footprint} is both sufficient and necessary for the complete report to survive a width extension. Theorem~\ref{thm:report-frame} uses two properties of the comparator. A successful proper expansion stays proper under width extension at a fixed required type, and the report records the endpoints at every stop. With fresh labels, inserting a field at any retained path changes the report while every other retained test passes. Corollary~\ref{cor:occurrence-impact} uses the paths for each occurrence to identify its changed report.

\begin{table}[t]
\centering\small
\caption{Guarantees for evidence reuse after edits}\label{tab:evidence-after-edits}
\begin{tabularx}{\linewidth}{@{}>{\raggedright\arraybackslash}p{.19\linewidth} *{3}{>{\raggedright\arraybackslash}X}@{}}
\toprule
Approach & Evidence & Edits considered & Established guarantee\\
\midrule
Nominal Adapton \citep{hammer-names-2015}
& Named dependency graph with cached results & Changes to mutable inputs & Updated result agrees with a fresh run on the new input\\
\emph{Provenance Traces} \citep[Theorem~10]{cheney-provenance-traces-2008}
& Query execution trace & Changes to the input store for the same query & Successful adaptation corresponds to traced evaluation on the new input\\
Certifying incremental type checking \citep{puech-certifying-2012}
& Checked typing subderivations & Program deltas referring to checked subderivations & Valid typing derivation for the edited program\\
Dependency provenance \citep{cheney-dependency-2011}
& Components of a query result & Changes to annotated inputs & Dependency-correct annotations for observed outputs\\
Dynamic slicing \citep{perera-explain-2012}
& Selected part of an execution result & Partial program and inputs ordered by filling holes & Least slice sufficient to explain the selected output\\
Trace and query slicing \citep[Theorem~4.4, Corollary~5.2]{cheney-query-slicing-2014}
& Output pattern, input slice and trace or query slice & Inputs agreeing on the slice, with a matching trace or a query extending the query slice & Successful recomputation preserves the selected output pattern\\
XML update independence \citep{benedikt-cheney-destabilizers-2010}
& Query result & Updates to an XML document & Disjointness from a sound destabilizer establishes independence\\
Identical-row elimination (Dedup, Section~\ref{sec:footprint-cost})
& Symbolic lower term, cut paths and endpoints & Fixed reads and validated width extension & One check per distinct local row; nested input paths can remain\\
Report reconstruction (Rebuild)
& Saved canonical report & Changes yielding supported B-valid certificates for the same export & Report identity decided by reconstruction and complete comparison\\
GCP (Theorem~\ref{thm:report-frame}, Corollaries~\ref{cor:source-footprint} and~\ref{cor:occurrence-impact})
& Source comparisons and synthesized capture footprint & Capture-type width extension and fresh projection for the same export & Exact report identity and affected occurrences; unique irredundant old-path tests at fixed reads with fresh labels\\
GCP (Corollary~\ref{cor:surplus})
& Old comparison stops and expanding records with equal required children & Two successful comparisons at one source occurrence in the ground record fragment of Section~\ref{sec:report-frame} & Exact local report identity, including deletions and read reselection; each surplus-field test has a deletion witness at fixed required type\\
\bottomrule
\end{tabularx}
\end{table}

\paragraph{Reusing a typing derivation.}
Puech and R\'egis-Gianas graft checked subderivations for retained subterms into a certificate for the edited program \citep{puech-certifying-2012}. Their example can omit an explicit subtyping step when the compared types coincide. GCP makes that coincidence observable when it constitutes the reviewed claim. A fresh projection can select different read bindings while the old typing witness remains valid. Theorem~\ref{thm:report-frame} identifies exactly when the reports of those two valid witnesses agree. Incremental typing and build systems develop algorithms for obtaining updated results and proofs efficiently \citep{wachsmuth-task-engine-2013,erdweg-cocontextual-2015,pacak-incremental-2020,mokhov-build-2018}.

\paragraph{Two separations on one mapper.}
Consider \texttt{r -> r.k.v}, with candidate types \(\{v:Int,flag:Bool\}\) and \(\{v:Int\}\), in that order, at the inner read and \(Int\) at the outer read. One fixed export admits all the following captures. Every execution reads \(k\) and \(v\) and returns 7.

First add \texttt{note=true} at the input root. Starting from \texttt{\{k=\{v=7\}\}}, the inner comparison originally stopped at root equality. It now follows \(k\), so its report changes. Starting from \texttt{\{k=\{v=7\},spare=true\}}, the inner comparison already followed \(k\). The same addition preserves both occurrences. A decision based on the executed reads, their values and the site of the addition receives identical observations in these two cases. Their old comparison stops distinguish them.

Now start from \texttt{\{k=\{v=7,spare=true\}\}} and add \texttt{flag=true} inside \(k\). At the old selection \(\{v:Int\}\), the inner comparison expands down to the primitive field \(v\), and the outer comparison depends on that selected type. The old source footprint is empty. Keeping the old read choices therefore preserves the report. Fresh projection can now select the first candidate, \(\{v:Int,flag:Bool\}\), and does so under this catalogue. The right endpoint changes at the inner read and the outer comparison also changes. Both certificates pass B. Even an exact test of every old capture subtree required at fixed read choices must thus be paired with the read-agreement test when certificates are projected independently.

\paragraph{One approval and three comparison procedures.}
Fix the whole-report approval and the two proposed insertions at \(r.k\) and \(r.k.v\) in Section~\ref{sec:footprint-cost}. An approval of one named equality covers only that equality. A recursive schema diff identifies both insertion sites. Rebuilding both reports and comparing leaves by occurrence gives the exact affected set after projection. With the old export, capture types and fixed read bindings, the source paths give that same set from the proposed sites. The shallower insertion changes one comparison, and the deeper insertion changes both. Thus the benefit before projection is knowing which published assertions would change. After projection the receiver can also compare the actual reports directly. Table~\ref{tab:rush-review} applies the member and source tests to the same published edits and checks both against reconstruction.

\paragraph{Dependency precision and least slices.}
\citet{cheney-dependency-2011} give a semantic account of dependency provenance, show that minimal dependency provenance is not computable for their query language, and provide conservative approximations. \citet[Section~3.4]{perera-explain-2012} establish least program/environment slices for selected partial outputs of a terminating run. Their ordering replaces parts of a fixed input and program by holes, and their evaluation semantics induces a Galois connection.

\citet{benedikt-cheney-schema-2009} use an XML schema to establish when an update preserves a query result. Their later analysis derives a \emph{destabilizer} from the query and tests an update for overlap with this approximation of the updates that can change the result \citep{benedikt-cheney-destabilizers-2010}. For a fixed capture and read assignment, view the old report as the query result and old-record insertion sites as the update vocabulary. The union of the descendant cones rooted at \(F_E\) is then an exact destabilizer in this vocabulary. Every batch meeting it changes the report, and every disjoint batch preserves it (Corollary~\ref{cor:source-footprint}). If each old record admits a fresh field, inserting one at a retained path changes the report while satisfying every other retained test. Over all width extensions, these witnesses establish the uniqueness of the irredundant conjunction of subtree equalities at old input-record paths.

%% file: figures/fig-09-unread-field-report.tex
\begin{figure}[tbp]
\centering
\begin{tikzpicture}[
  font=\small, >={Stealth[length=2mm]}, line width=.5pt,
  comparison/.style={draw=black!40, fill=black!3, rounded corners=2pt,
    align=center, text width=6.45cm, minimum height=.79cm, inner sep=3pt},
  leaf/.style={comparison, minimum height=1.02cm}
]
\node[font=\small\bfseries] at (0,0) {Program \texttt{r -> r.k} \enspace \textbar{} \enspace selected read type \(Int\) \enspace \textbar{} \enspace return value \(7\)};
\node[font=\small\bfseries] at (-3.58,-.61) {Before the extension};
\node[font=\small\bfseries] at (3.58,-.61) {After adding unread field \texttt{note}};
\node[comparison] (before) at (-3.58,-1.24) {\(\{k:Int\}\preceq_K\{k:Int\}\)};
\node[comparison] (after) at (3.58,-1.24) {\(\{k:Int,note:Bool\}\preceq_K\{k:Int\}\)};
\node[leaf, fill=white, draw=black] (rootleaf) at (-3.58,-2.95)
  {Root leaf \enspace \(p=\epsilon\)\\[2pt]\((\{k:Int\},\{k:Int\})\)};
\node[leaf, fill=black!10, draw=black] (fieldleaf) at (3.58,-2.95)
  {Field leaf \enspace \(p=k\)\\[2pt]\((Int,Int)\)};
\draw[->] (before) -- node[right, font=\footnotesize] {Records equal} (rootleaf);
\draw[->] (after) -- node[right, font=\footnotesize] {Follow required field \(k\)} (fieldleaf);
\end{tikzpicture}
\caption{How an unread field changes a comparison report}
\Description{Two panels compare report decomposition for the same field read. The first stops at the root because the records are equal. The second adds an unread field, follows k, and stops at the comparison of the two Int types.}
\label{fig:unread-field-report}
\end{figure}

%% file: figures/fig-10-report-preservation-test.tex
\begin{figure}[tbp]
\centering
\begin{tikzpicture}[
  font=\small, >={Stealth[length=2mm]}, line width=.5pt,
  test/.style={draw=black!40, fill=black!3, rounded corners=2pt,
    align=center, text width=3.04cm, minimum height=1.01cm, inner sep=3pt},
  result/.style={test, text width=2.86cm, minimum height=.77cm}
]
\node[align=center,font=\small\bfseries] at (0,0) {Same Outline\(_R\) export and two completed B-valid certificates\\Ground types in the theorem's fragment and width-extended captures};
\node[test] (read) at (-4.68,-1.24) {Same selected\\read bindings?};
\node[test] (cut) at (0,-1.24) {Same left subtrees\\at all old cuts?};
\node[result, fill=white, draw=black] (same) at (4.68,-1.24) {Report preserved};
\node[result, fill=black!10, draw=black] (changed) at (-2.34,-2.89) {Report changed};
\draw[->] (read) -- node[above, font=\footnotesize] {yes} (cut);
\draw[->] (cut) -- node[above, font=\footnotesize] {yes} (same);
\draw[->] (read.south) -- node[left, font=\footnotesize] {no} (-4.68,-2.89) -- (changed.west);
\draw[->] (cut.south) -- node[right, font=\footnotesize] {no} (0,-2.89) -- (changed.east);
\end{tikzpicture}
\caption{The decision procedure for report preservation}
\Description{Flow chart under the theorem's premises. If the selected read bindings differ or a left subtree at an old cut differs, the report changed. If both checks pass, the report is preserved.}
\label{fig:report-preservation-test}
\end{figure}

%% file: sections/support/source-footprint.tex
\paragraph{Synthesizing a source-level footprint.}\label{sec:source-footprint}
Different obligations can observe the same capture, or protect a subtree already covered by another obligation. We compile their cuts into one set of capture checks. Fix an Outline\(_R\) export \(E\), an assignment \(\rho\) to all parameter variables and an assignment \(\beta\) to all read variables, with successful original comparisons in Theorem~\ref{thm:report-frame}'s ground fragment. Write \(\theta=\rho\uplus\beta\) and \(\mathcal R_E(\rho,\beta)\) for the complete canonical report in a fixed session.

For a symbolic lower type \(\tau\) and a path \(p\) defined in \(\theta\tau\), define a pullback \(\Phi(\tau,p)\) to parameter/path pairs. At \(Ref(x)\) for a parameter, it returns \(\{(x,p)\}\) if \(\rho(x)|_p\) is a record, and the empty set otherwise. At a read reference it returns the empty set, since \(\beta\) is fixed. At a record constructor and path \(kq\), it continues with the \(k\)-child at \(q\). At a record constructor and the empty path, it unions the pullbacks of all children at the empty path. A primitive, top or bottom leaf contributes the empty set. These rules follow the exported type expression, including references inside constructed records.

For the original roots \((o,L_o,U_o)\in O\), form
\[
 H_E=\bigcup_{o\in O}\ \bigcup_{p\in K_R(\theta L_o,\theta U_o)}
       \Phi(L_o,p),
 \qquad F_E=\operatorname{minpref}(H_E).
\]
Here \(\operatorname{minpref}\) removes duplicates and removes \((x,q)\) whenever a distinct \((x,p)\) has \(p\) as a prefix of \(q\). Paths belonging to different parameters remain separate.

\begin{gcorollary}[Exact source footprint]\label{cor:source-footprint}
Let \(\rho'\) range over all componentwise width extensions of \(\rho\) in the ground fragment, with \(\beta\) fixed. Then
\[
 \mathcal R_E(\rho,\beta)=\mathcal R_E(\rho',\beta)
 \quad\Longleftrightarrow\quad
 \forall(x,p)\in F_E.\ \rho(x)|_p=\rho'(x)|_p.
\]
If a fresh field label is available at every old record path, \(F_E\) is the unique irredundant set of equality tests at old capture-record paths whose conjunction characterizes report preservation over these extensions.
\end{gcorollary}
\begin{proof}
Structural induction on \(\tau\), following \(p\), equates preservation of \((\theta\tau)|_p\) with agreement on \(\Phi(\tau,p)\). Parameter references give the indicated capture subtree; non-record capture subtrees stay fixed under width extension. Read references and ground constants stay fixed. Record constructors preserve their field sets, so their equality reduces to equality of the corresponding children. Theorem~\ref{thm:report-frame} and Corollary~\ref{cor:record-cuts} therefore reduce report equality to agreement on \(H_E\). Agreement at an ancestor implies agreement at each descendant, giving the same predicate for \(F_E\).

For \((x,p)\in F_E\), add a fresh Boolean field at \(p\) in \(\rho(x)\). Prefix-freeness preserves every other frontier test while changing this one, so the report changes. Thus every retained test is necessary. For uniqueness, let \(G\) be any exact conjunction of old capture-record tests. The same insertion at \(p\) forces \(G\) to contain \((x,q)\) with \(q\) a prefix of \(p\). Inserting at \(q\) fails that test, so exactness forces some \((x,r)\in F_E\) with \(r\) a prefix of \(q\). Prefix-freeness of \(F_E\) gives \(r=q=p\). Hence \(F_E\subseteq G\). Every further test of \(G\) lies below a frontier path by the same insertion argument and is redundant. Therefore the unique irredundant conjunction is \(F_E\).
\end{proof}

\paragraph{Locating a batch of field additions.}
For a componentwise width extension \(\rho\mathrel{\unlhd}\rho'\), let \(I_x\) be the set of paths to records in the old capture \(\rho(x)\) that gain at least one immediate field. An inserted record contributes its attachment site in the old tree. Write \(p\sqsubseteq q\) for the prefix relation, including equality. For each original root \((o,L_o,U_o)\), retain its local links before reducing \(H_E\) to \(F_E\). With \(o\) ranging over original occurrence IDs, define
\[
\begin{aligned}
 W_o &= \{(k,x,p)\mid k\in K_R(\theta L_o,\theta U_o),\ (x,p)\in\Phi(L_o,k)\},\\
 A(I)&= \{o\mid \exists(k,x,p)\in W_o,\ \exists q\in I_x:\ p\sqsubseteq q\}.
\end{aligned}
\]
Here \(k\) identifies a stop in the old comparison, \(p\) a record in its input, and \(q\) the record where a field was attached. Let \(\mathcal R_o\) be the ordered leaves of occurrence \(o\) in the fixed session.

\begin{gcorollary}[Exact impact at source occurrences]\label{cor:occurrence-impact}
Fix the export, successful old comparisons and read assignment \(\beta\) of Corollary~\ref{cor:source-footprint}. For every componentwise width extension in the ground fragment, with insertion sites \(I\) as above,
\[
 o\in A(I)\quad\Longleftrightarrow\quad
 \mathcal R_o(\rho,\beta)\ne\mathcal R_o(\rho',\beta).
\]
The complete report is preserved exactly when \(A(I)\) is empty. Equivalently, it is preserved exactly when there are no \((x,p)\in F_E\) and \(q\in I_x\) with \(p\sqsubseteq q\).
\end{gcorollary}
\begin{proof}
Apply Theorem~\ref{thm:report-frame} to the root at \(o\). Its report changes exactly when the left subtree at some old cut changes. Non-record cuts stay fixed under width extension. For a record cut \(k\), the pullback argument of Corollary~\ref{cor:source-footprint} equates preservation of that subtree with equality at every input record in \(\Phi(L_o,k)\). Such a record at \((x,p)\) changes precisely when a field is added at \(p\) or at an old descendant \(q\). These are exactly the witnesses in \(A(I)\), for any number of insertions. The complete report concatenates the ordered leaves of the distinct original occurrences. Taking the union of their tests therefore decides equality of the complete report. Removing a descendant test from this union preserves the decision because every hit on it also hits its ancestor.
\end{proof}
\paragraph{When global tests identify the affected occurrences.}
Fix the export, old captures and read bindings of Corollary~\ref{cor:occurrence-impact}. Let \(F_o=\operatorname{minpref}\{(x,p)\mid \exists k.\,(k,x,p)\in W_o\}\) be the paths retained for occurrence \(o\) alone. For insertion sites \(I\), write \(b_E(I)\) for the Boolean vector recording which subtrees at paths in \(F_E\) change. Assume a fresh field name is available at every old record. Then
\begin{equation}\label{eq:attribution-from-global}
  \bigl(\forall I,J.\ b_E(I)=b_E(J)\ \Longrightarrow\ A(I)=A(J)\bigr)
  \quad\Longleftrightarrow\quad
  \forall o.\ F_o\subseteq F_E .
\end{equation}
Here \(I,J\) range over insertion sites of all width extensions of the old captures. The left side says that two edits with identical global test outcomes change exactly the same source comparisons. The right side says that every path needed by any one comparison survives global reduction.

To see sufficiency, apply Corollary~\ref{cor:occurrence-impact} at \(o\). Removing descendant paths within that occurrence leaves its decision unchanged. The occurrence changes exactly when a test at one of its paths in \(F_o\) fails, and each such outcome is in the vector. For necessity, suppose a local path \(q\in F_o\) is absent from \(F_E\), writing paths relative to their capture. Global reduction retained a strict ancestor \(p\) of \(q\) in that capture. Compare two edits to the old input, adding a fresh Boolean field directly at \(p\) or directly at \(q\). Both fail only the global test at \(p\), since \(F_E\) is prefix-free within each capture. The insertion at \(q\) changes occurrence \(o\). The insertion at \(p\) misses every path in \(F_o\), since any local path covering \(p\) would also strictly cover \(q\) and preclude its retention. Thus the same vector gives different affected sets. For a supplied edit, the occurrence links in Corollary~\ref{cor:occurrence-impact} decide the affected set without the fresh-label assumption.

Using source parameter names for the footprint, consider \texttt{r -> \{a=r.k,b=r.k\}} with \texttt{r} of type
\(\{k:\{v:Int\},spare:Bool\}\) and \(\{v:Int\}\) selected at both reads. The two roots each protect \(k\); synthesis yields the single test \((r,k)\). Adding a field beside \(k\) preserves both reports, while adding one inside \(k\) changes them. For \texttt{r -> \{box=\{left=r,right=r\}\}.box}, a whole-record equality at the constructed host pulls back to one test of \(r\), despite its two occurrences.

The uniqueness statement ranges over all componentwise width extensions. A deployment catalogue can restrict those extensions, for example by admitting only captures with the same type at a particular path. The receiver first compares the selected read bindings and then checks \(F_E\) directly on captures. This compiles the per-root characterization into a shared source-level condition while retaining the separate occurrence identities in the report. Catalogue admission and output-shape validation compose with that condition in the delivery contracts of Section~\ref{sec:consumer-contract}.

%% file: sections/support/finite-kernel.tex
\paragraph{The kernel behind the source contract.} The interface \(T_\rho(E)\) above uses explicit ground candidates and the executable relation below. Its certificate checker supplies B with the same relation used by source export. Table~\ref{tab:claim-correspondence} fixes the scope of the wider carrier adapter.

\subsection{An Executable Template Calculus}\label{sec:calculus}

\subsubsection{Types and well-formed templates}
We use finite, acyclic type trees. Let $b$ range over ground atoms, $v$ over $V$, and $\ell$ over distinct field labels. The syntax is
\[
\begin{array}{rcl}
\tau&::=&b\mid\bot\mid\top\mid v\mid\{\ell_i:\tau_i\}_{i\in I}
 \mid\tau\to\tau\mid\mathsf{Array}(\tau)\\
&&{}\mid\mathsf{Dict}(\tau,\tau)\mid\mathsf{Option}(\tau_1,\ldots,\tau_k).
\end{array}
\]
An $\mathsf{Array}$ here is immutable. A $\mathsf{Dict}$ describes storage that may be mutated and therefore has invariant parameters. The least type $\bot$ and the inhabited atom $\mathsf{Unit}$ are distinct.

The supplied atoms include $\mathsf{Int}$, $\mathsf{Long}$, $\mathsf{Number}$, $\mathsf{Float}$, $\mathsf{Decimal}$, $\mathsf{Double}$, $\mathsf{String}$, $\mathsf{Bool}$, and $\mathsf{Unit}$. The nontrivial primitive order is generated by
\[
\mathsf{Int}\preceq_K\mathsf{Long}\preceq_K\mathsf{Number},\quad
\mathsf{Float}\preceq_K\mathsf{Decimal}\preceq_K\mathsf{Number},\quad
\mathsf{Double}\preceq_K\mathsf{Number}.
\]
These primitive edges define the abstract compatibility relation used by the kernel. Field labels use case-sensitive string equality and lexicographic UTF-16 code-unit order, as in Java's string comparison. Public constructors copy fields into this canonical order, independently of any comparator supplied by a host-language map.

An Option denotes a finite union for compatibility checking. Every public type constructor normalizes its immutable children and flattens nested Options, removes $\bot$, deduplicates structurally equal arms, and orders the remaining arms deterministically. An empty union becomes $\bot$ and a singleton becomes its element. Distinct arms remain present even when one is covered by another. Structural equality of the resulting trees, written $=$, is consequently stronger than mutual subtyping. This distinction is used by the exact-match selector rule. Java centralizes normalization in the \texttt{Type} constructor, including calls from \texttt{of}, \texttt{option} and the binary decoder. Consequently candidates, bounds, selectors, actuals and returns share the same recursive invariant before they enter a template or certificate. Option arms are deduplicated by structural equality; their deterministic order uses display text in the same UTF-16 order followed by a structural tie-break, so punctuation in a field label cannot merge different records. Python value typing, substitution and binary decoding use this common canonical order. Its text reader consumes every character, checks constructor arity and rejects duplicate field labels before normalization. Text observations use ASCII identifiers for atoms, variables and field labels. In-memory record types admit arbitrary Java strings as field labels. The binary format represents labels that can be encoded as well-formed UTF-8.

A ground substitution $\theta$ maps every $v\in V$ to a ground type. Simultaneous substitution descends into all constructors, including both sides of arrows, dictionary keys and values, fields, Option arms, and return types. It normalizes each reconstructed Option. No free reference may remain after substitution. Since substituted values are ground, substitution follows no variable-to-variable chain and cannot create a recursive type.

\begin{gdefinition}[Template]\label{def:template}
A template is $T=(V,\mathcal D,C,\Sigma,p,r)$, where $\mathcal D(v)$ is a finite nonempty ordered list of structurally distinct ground candidates for $v$, $C$ is an ordered list of tagged bounds $(L,U,s,\lambda)$, $\Sigma$ is an ordered list of selectors, and $p,r$ are the formal and return type expressions. Every reference in $C,\Sigma,p,r$ belongs to $V$. Lists and types are immutable. Candidate order is part of the input to deterministic search.
\end{gdefinition}
After type normalization, Java's \texttt{Template} constructor removes structurally equal duplicate candidates, preserving their first occurrence and the variable insertion order. Definition~\ref{def:template} and Algorithm~\ref{alg:projection} operate on this resulting template. Mutually compatible but structurally different candidates remain distinct. Search counts and budget outcomes therefore refer to the deduplicated lists.
For the source instance, Section~\ref{sec:source-residual} constructs \(T_\rho(E)\) with the uncaptured domains, all partially substituted roots, empty selectors and Unit actual/formal. Root metadata remains available outside the search tuple for final-evidence auditing. The domain restriction is explicit. A finite-height order can have infinitely many incomparable elements; finite height alone would not justify exhaustive search. Here the number of candidates is supplied and measurable. The adapter constructs its candidate lists from actual types, exported bounds and fixed primitive candidates.

\subsubsection{Ground structural compatibility}\label{sec:subtype}
The ground judgment $A\preceq_K B$ is the following structural decision procedure. Equality, $\bot\preceq_K B$, and $A\preceq_K\top$ succeed first. Remaining clauses are considered in the listed order.
\begin{enumerate}
\item If $A=\mathsf{Option}(A_1,\ldots,A_k)$, require $A_i\preceq_K B$ for every arm. This precedes the right-Option clause.
\item Otherwise, if $B=\mathsf{Option}(B_1,\ldots,B_m)$, require $A\preceq_K B_j$ for some arm.
\item For records, each label in $B$ must also occur in $A$, and the field type in $A$ must be a subtype of the corresponding field type in $B$.
\item For arrows, $A_1\to A_2\preceq_K B_1\to B_2$ exactly when $B_1\preceq_K A_1$ and $A_2\preceq_K B_2$.
\item For immutable arrays, compare element types covariantly.
\item For dictionaries, require mutual subtyping of the key types and mutual subtyping of the value types.
\item For atoms, use the primitive order above. All remaining cases fail.
\end{enumerate}
For example, $(\mathsf{Number}\to\mathsf{Int})\preceq_K(\mathsf{Int}\to\mathsf{Number})$ succeeds; the reverse fails. Likewise, $\mathsf{Dict}(\mathsf{String},\mathsf{Int})$ is not below $\mathsf{Dict}(\mathsf{String},\mathsf{Number})$: the latter would permit writes that violate the former's storage contract.

These clauses define the structural relation used by the theorems. Checking whether a non-Option type is below a union requires a single covering arm.

\begin{glemma}[Ground decision]\label{lem:ground}
Ground compatibility terminates and decides the clauses above. It is reflexive and transitive on the finite type syntax.
\end{glemma}
\begin{proof}
Structural recursion decreases total operand size. Reflexivity uses equality; transitivity composes record inclusion, reversed arrow domains, covariant array elements, invariant dictionary parameters and the ordered Option cases. Section~\ref{app:finite-kernel} gives the constructor induction.
\end{proof}

\subsubsection{Closed Option selection}\label{sec:selector}
Branch selection adds an admissibility policy to union compatibility. A selector is $\sigma=(a,[b_1,\ldots,b_k],\pi,\lambda)$ with polarity $\pi\in\{+,-\}$. Substitute $a$ and all arms, then deduplicate structurally equal ground arms, retaining their first indices. At positive polarity a covering arm satisfies $a\theta\preceq_K b_i\theta$; at negative polarity it satisfies $b_i\theta\preceq_K a\theta$. Selection succeeds when there is exactly one covering arm, or when exactly one of several covering arms is structurally equal to $a\theta$. Otherwise it fails.

This is an additional admissibility policy. For example, $\{n:\mathsf{Int},t:\mathsf{String}\}$ is covered by both $\{n:\mathsf{Number}\}$ and $\{t:\mathsf{String}\}$, but neither is exact, so selection fails.

Selection must follow substitution. Two symbolic arms can become identical ground types; counting them twice creates spurious ambiguity. Treating an unbound arm variable as top can select a branch that stops covering the argument after specialization. The kernel tests a selector only when all references are bound, and its final checker repeats the test on the total substitution. Considering an arm never mutates the shared template.

\noindent\begin{minipage}{\linewidth}
\subsubsection{Admissible certificates}\label{sec:certificate}
\begin{gdefinition}[Successful projection]\label{def:valid}
For ground actual $A$, a certificate $(\theta,R)$ is valid, written $T;A\vdash\theta\Rightarrow R$, exactly when
\begin{align}
\operatorname{dom}(\theta)&=V,&\theta(v)&\in\mathcal D(v)&&(v\in V),\label{eq:domain}\\
A&\preceq_K p\theta,\label{eq:actual}\\
L\theta&\preceq_K U\theta&&&((L,U,s,\lambda)\in C),\label{eq:bounds}\\
\operatorname{select}(\sigma,\theta)&\text{ succeeds}&&&(\sigma\in\Sigma),\label{eq:select}\\
R&=r\theta.\label{eq:return}
\end{align}
\end{gdefinition}
\end{minipage}\par
These equations define the entire finite-template certificate. The additional source obligation is discharged for Outline\(_R\) in Section~\ref{sec:source-residual}.

Every original bound is checked after simultaneous substitution. Consider $\mathcal D(x)=[\mathsf{Int},\mathsf{Number}]$, $\mathcal D(y)=[\mathsf{Number}]$, edge $y\preceq_K x$, formal and return $x$, and actual $\mathsf{Int}$. Choosing $x=\mathsf{Int}$ first looks compatible with the call but violates the edge once $y=\mathsf{Number}$. The valid solution is $x=y=\mathsf{Number}$. Committing to the first local choice and then rejecting would be incomplete even for these finite domains. The delivered search retains and explores the alternative.

\subsection{Projection Algorithm and Guarantees}\label{sec:algorithm}

\subsubsection{Search and independent checking}
Algorithm~\ref{alg:projection} uses finite-domain backtracking with forward checking. Mutable state consists of a partial substitution, counters, and a diagnostic string, all owned by one invocation. The actual/formal edge is added to a session-local copy of the bounds. A partial state is consistent when every fully instantiated bound and selector succeeds. Obligations with unbound references remain pending until those references are bound.

\begin{figure}[htbp]
\refstepcounter{gcpalgorithm}\label{alg:projection}
\small
\begin{tabularx}{\linewidth}{rX}
\toprule
1 & $\operatorname{Project}(T,A,B)$: allocate fresh state and append $(A,p,\mathsf{CALL})$ to a local bound list.\\
2 & Run $\operatorname{Search}(\varnothing)$. Before each ground obligation check, consume one unit of budget $B$.\\
3 & $\operatorname{Search}(\eta)$: if an instantiated obligation fails, fail this branch.\\
4 & If $\operatorname{dom}(\eta)=V$, return a copy of $\eta$.\\
5 & For every unbound $v$, try each $t\in\mathcal D(v)$ temporarily. Retain $t$ exactly when $\eta[v\mapsto t]$ is consistent. Restore $\eta$ after every trial.\\
6 & If any variable has no retained value, fail the branch. Otherwise choose a variable with the fewest retained values, breaking ties by input order.\\
7 & Recurse on its retained values in input order, restoring the binding afterward. Return the first successful total substitution; fail if all branches fail.\\
8 & Exhausted search returns \textsc{NoSolutionInDomain}; exhausted budget returns \textsc{Inconclusive}.\\
9 & On success, compute $R=r\theta$ and check the original $T,A,\theta,R$ with the separate certificate checker. Return \textsc{Success} only if it accepts.\\
\bottomrule
\end{tabularx}
\par\smallskip\centering\textbf{Algorithm \thegcpalgorithm.} Checked finite-domain projection. The checker enumerates every original obligation.\par
\Description{Nine numbered steps specify fresh allocation, closed-obligation checking, forward checking, finite search, scoped failure, and final certificate validation.}
\end{figure}

The checker independently traverses the immutable inputs. It verifies domain membership, the actual/formal edge, every original bound, every ground selector, and the substituted return. Search and checking share the ground compatibility implementation.

The implementation increments its check counter immediately before testing the budget. An exhausted run records $B+1$ attempted checks, of which at most $B$ were executed. A selector check can compare many arms, and a structural comparison can traverse many nodes. The budget is denominated in obligation checks; selector width and structural depth determine the work per check.

\subsubsection{Validity and relative completeness}
\begin{glemma}[Extension-invariant failure]\label{lem:pruning}
If all variables of a bound or selector are bound by a partial substitution $\eta$ and it fails under $\eta$, no extension of $\eta$ can satisfy it.
\end{glemma}
\begin{proof}
An extension changes only variables outside $\operatorname{dom}(\eta)$. None occurs in the tested obligation. Substitution therefore produces the same ground types and the same deterministic selection outcome under every extension.
\end{proof}

\begin{gtheorem}[Projection validity]\label{thm:valid}
If projection returns \textsc{Success}$(\theta,R)$, then $T;A\vdash\theta\Rightarrow R$.
\end{gtheorem}
\begin{proof}
The final checker directly conjoins Equations~\eqref{eq:domain}--\eqref{eq:return}. The ground procedure terminates and implements the specified relation by Lemma~\ref{lem:ground}. Search returns success only after this conjunction is true. Every original bound and the return use the same total substitution; independently chosen bindings cannot evade a cross-variable obligation.
\end{proof}

\begin{gtheorem}[Termination and finite-domain completeness]\label{thm:complete}
Without a budget cutoff, search terminates and succeeds if and only if a valid certificate exists in $\prod_{v\in V}\mathcal D(v)$. With a finite budget it terminates with \textsc{Success}, \textsc{NoSolutionInDomain}, or \textsc{Inconclusive}. The last outcome makes no satisfiability claim.
\end{gtheorem}
\begin{proof}
Each recursive edge binds one previously unbound variable, so depth is at most $|V|$. Each domain is finite, giving finite branching. Consistency tests terminate by structural recursion and finite selector enumeration. Thus the unbounded search tree is finite.

Suppose a valid total substitution $\theta^*$ extends the current partial state. For every unbound variable $v$, the trial value $\theta^*(v)$ cannot be pruned: otherwise Lemma~\ref{lem:pruning} contradicts validity of $\theta^*$. No domain is empty, and whichever variable the heuristic chooses retains its value from $\theta^*$. Repetition supplies a successful branch of the finite tree. Search reaches that branch or succeeds earlier with another valid certificate. The converse is Theorem~\ref{thm:valid}. A cutoff interrupts enumeration, so the result must remain inconclusive.
\end{proof}
Candidate coverage and selection order determine the result. An identity template with actual $\mathsf{String}$ and domain $\mathcal D(x)=[\mathsf{Int}]$ has no solution, despite the useful omitted candidate $\mathsf{String}$. Reversing $[\mathsf{Int},\mathsf{Number}]$ can change an identity projection's result for actual $\mathsf{Int}$, although both results are valid. Minimum-remaining-values ordering controls the search; the returned certificate also depends on the supplied domain order.

\subsubsection{Freshness and source erasure}
\begin{gtheorem}[Template invariance and independent calls]\label{thm:fresh}
Completed, rejected, and budget-interrupted calls leave a template's domains, bounds, selectors, formal, and return unchanged. Reordering calls leaves each call's result unchanged when its input, domain order, and budget are fixed.
\end{gtheorem}
\begin{proof}
The algorithm reads immutable template components and allocates mutable state per invocation. The added call bound belongs to a copied list; writes affect only the local partial substitution. Returned substitutions are copied. A budget exception discards local state. No call changes another call's input or search order, so deterministic search gives the same per-call result after any reordering.
\end{proof}
For partial application, the source layer supplies the residual template and captured environment (Section~\ref{sec:source-residual}); the finite kernel specializes arrow-shaped types within this interface.

\begin{gtheorem}[Source erasure]\label{thm:erasure}
Replacing every bound's source label by $\mathsf{ERASED}$ preserves valid certificates. With identical list order and budget it preserves search status, substitution, result, and operation counts. A diagnostic source string may change.
\end{gtheorem}
\begin{proof}
None of Equations~\eqref{eq:domain}--\eqref{eq:return} inspects a source label. Search chooses variables by domain order and viable-candidate count; consistency depends on the same ground bounds and selectors. Induction over search steps gives identical states, decisions, and counters. A source field is read only when recording a failed bound's diagnostic.
\end{proof}
Provenance retains origin for inspection and lifecycle operations. Semantic acceptance depends on the associated constraints. The erasure theorem makes the distinction precise.

\subsubsection{Complexity in terms of supplied input}\label{sec:complexity}
The bounds start from the normalized immutable template. Let $n=|V|$, $d=\max(1,\max_{v\in V}|\mathcal D(v)|)$ (taking $d=1$ when $V$ is empty), and $c=|C|+|\Sigma|+1$, including the call edge. Let $S$ bound all supplied type expressions and candidates, and the ground expansions of obligations and the return before and after normalization, including all selector arms. Let $K(S)$ bound the cost of finding references, substitution and normalization, comparisons, equality/hashing, and selector enumeration. It includes Java tree-representation costs and is not treated as a constant.

There are at most $\sum_{i=0}^{n}d^i$ search nodes. At each node, forward checking tries at most $nd$ values and each consistency scan considers at most $c$ obligations. An input-parameter bound is therefore
\[
 O\!\left(\left(\sum_{i=0}^{n}d^i\right)(1+nd)cK(S)\right),
\]
plus one final certificate check on success. For $d=1$ the sum is $n+1$; for $d\ge2$ it is $O(d^n)$. Search is exponential in the number of variables in the worst case. Ground compatibility visits finitely many subterm pairs, but hashing, equality, and substitution costs still belong in $K(S)$. These representation operations determine the cost per visited pair.

There are at most $n$ stack levels, each retaining at most $d$ candidate references for its chosen variable. Persistent search bookkeeping is $O(nd+n)$ beyond the immutable input and temporary comparison/substitution data. Substitution can duplicate candidate trees; $S$ includes that expansion. The candidate pool stays fixed throughout search.

\subsubsection{Proof Details for the Executable Finite Kernel}\label{app:finite-kernel}

\paragraph{Ground substitution}
Let $\operatorname{fv}(\tau)$ collect references structurally: it is empty for atoms, $\{v\}$ for a reference, the union of the children's sets for constructors, and the union over field types for a record. If $\operatorname{fv}(\tau)\subseteq\operatorname{dom}(\eta)$ and every image of $\eta$ is ground, then $\tau\eta$ is ground. The proof is structural induction. The reference case returns the ground image; every constructor combines ground children. Option normalization only removes, sorts, or flattens existing ground arms and cannot introduce a reference.

If $\eta'$ extends $\eta$ without changing existing bindings, then $\tau\eta'=\tau\eta$ whenever all free references of $\tau$ are in $\operatorname{dom}(\eta)$. Induct on $\tau$ again. The only interesting case is a reference, whose image is fixed by extension. For Options, the input arm sequences are identical after substitution, so deterministic normalization gives identical outputs. Applying this result to every type expression in a bound or selector proves Lemma~\ref{lem:pruning}, including symbolic-arm deduplication.

\paragraph{Transitivity of the ground relation}
For completeness, consider $A\preceq_K B$ and $B\preceq_K C$. Equalities allow immediate substitution in the other premise. Bottom on the left and top on the right are immediate. Other top/bottom cases reduce to equality or to the Option clauses containing those types. For the remaining cases, induct on the sum of the sizes of $A,B,C$.

If $A$ is an Option, its left-Option rule supplies $A_i\preceq_K B$ for every arm. The induction hypothesis supplies $A_i\preceq_K C$; the left-Option rule yields the conclusion. Otherwise, if $B$ is an Option, the first premise chooses an arm $B_j$ with $A\preceq_K B_j$. The second premise, using its left-Option rule, supplies $B_j\preceq_K C$. Induction gives $A\preceq_K C$. Otherwise, if $C$ is an Option, the second premise chooses $C_k$ with $B\preceq_K C_k$. Induction gives $A\preceq_K C_k$, and the right-Option rule gives $A\preceq_K C$.

After these cases, all three types have non-Option outer constructors. Nontrivial compatibility requires matching structural constructors or primitive-order edges. For records, a label required by $C$ occurs in $B$ and hence in $A$. The two premises give $A(\ell)\preceq_K B(\ell)\preceq_K C(\ell)$, to which induction applies. Thus the fields required by $C$ exist in $A$ with compatible types.

For arrows, write $A=A_d\to A_r$, and similarly for $B,C$. The premises give $B_d\preceq_K A_d$ and $C_d\preceq_K B_d$, so induction yields $C_d\preceq_K A_d$. On returns, $A_r\preceq_K B_r\preceq_K C_r$ yields $A_r\preceq_K C_r$. These are exactly the arrow-rule premises for $A\preceq_K C$. The domain reversal is essential.

For arrays, compose the element judgments. For dictionaries, the premises give compatibility in both directions for each parameter. For keys, $A_k\preceq_K B_k\preceq_K C_k$ and $C_k\preceq_K B_k\preceq_K A_k$ yield mutual compatibility of $A_k,C_k$; repeat for values. Primitive edges compose because the supplied primitive relation includes their transitive consequences. This exhausts the successful clauses of the decision procedure.

The relation is a preorder. For example, \(\mathsf{Option}(\mathsf{Int},\mathsf{Number})\) and \(\mathsf{Number}\) are mutually compatible and have distinct normalized trees. A selector tests for an exact match by comparing those trees.

\paragraph{Search completeness under dynamic variable ordering}
The forward-checking heuristic recomputes viable values at every node and may change which variable it selects as bindings accumulate. Completeness does not depend on a fixed variable order. For a particular valid $\theta^*$ extending the current state, each unbound variable retains at least its image under $\theta^*$. Therefore whichever variable is chosen has a viable child consistent with $\theta^*$. The number of unbound variables decreases, allowing induction on that number. Search may visit other children first, but the entire tree is finite. No rule prunes a value merely because a speculative bound with an unbound reference would fail.

A total state is checked again by consistency before it is returned. The outer checker additionally validates exact domain membership and the result equation. Thus a future change to search that forgets a bound or mishandles return construction can be detected by the checker, provided the checker's relation and template input remain correct. Sharing the same ground relation still leaves a common dependency; this motivates the independent Python implementation.

\paragraph{Candidate order and domain-relative completeness}
Consider a domain $[\mathsf{Number},\mathsf{Int}]$, no definition bounds, formal $x$, return $x$, and actual $\mathsf{Int}$. Both assignments are valid, and deterministic search can return $\mathsf{Number}$ first. The search order selects a witness from the supplied product. Relative completeness ensures that search finds a witness whenever that product contains one and the budget is unrestricted.

Exhaustive failure establishes that the supplied product contains no solution. A caller can widen the candidate domain and obtain a solution. Widening a domain preserves existing valid certificates but can change which one is returned, because it changes order and the minimum-remaining-values heuristic. If a budget is active, it can also change whether search reaches an existing solution before cutoff. The receiver reports domain exhaustion and budget exhaustion as distinct outcomes.

\paragraph{Source erasure and order}
Tag erasure preserves the order and multiplicity of type obligations. This matters for exact operational agreement: permuting semantically equivalent constraints can change early failure points and operation counts, particularly with a budget. Theorem~\ref{thm:erasure} holds list order fixed. Theorem~\ref{thm:fresh} concerns the order of separate calls, not arbitrary reordering of constraints within one call. In both cases, diagnostics are distinguished from semantic certificates.

\subsection{Implementation Correspondence}\label{sec:bridge}

\subsubsection{The executable Java kernel}
The implementation is in package \path{org.twelve.gcp.kernel}. Its classes are \texttt{FiniteProjection}, \texttt{Projection\allowbreak Certificate\allowbreak Checker}, and \texttt{Carrier\allowbreak KernelBridge}. The first two implement the interface specified above. Table~\ref{tab:correspondence} identifies the relevant representations. The recursive carrier implements a separate compatibility relation.

\begin{table}[t]
\caption{Finite kernel specification and implementation}
\label{tab:correspondence}
\small
\begin{tabularx}{\linewidth}{p{28mm}p{49mm}X}
\toprule
Formal object & Java operation / representation & Obligation\\
\midrule
Type tree & \texttt{Type}, \texttt{substitute} & Close references under every constructor; normalize rebuilt Options.\\
$A\preceq_K B$ & \texttt{subtype}, private \texttt{sub} & Ordered union cases, record fields, reversed arrow domain, invariant Dict.\\
Tagged bound & \texttt{Bound} & Retain original types, source, and location.\\
Ground domains & \texttt{Template.domains} & Ground nonempty lists; reject undeclared references.\\
Fresh search & \texttt{Search.solve}, \texttt{consistent} & Local bindings; only closed obligations prune.\\
Option selector & \texttt{selection} & Substitute, deduplicate, select unique cover or exact cover.\\
Validity judgment & \texttt{check} & Check domain, actual, all bounds and selectors, and substituted return.\\
Scoped outcome & \texttt{Status} & Distinguish domain exhaustion from budget exhaustion.\\
\bottomrule
\end{tabularx}
\end{table}

The Java records copy their input lists and maps. The type record stores an unmodifiable sorted field map; domains and returned substitutions are copied and wrapped. Under ordinary Java collection semantics, clients cannot mutate these components through their accessors. The representation map reads atom tags as atoms and reference tags as variables, recursively translates children and fields, and preserves template list order.

For this representation, substitution and compatibility follow the formal clauses by structural induction. \texttt{consistent} checks each obligation once its references are bound; temporary insertion/removal implements line~5 of Algorithm~\ref{alg:projection}. \texttt{solve} performs the stated recursion and tie-breaking. The checker implements Definition~\ref{def:valid}'s conjunction.

\paragraph{Development and verification.}
Section~\ref{sec:development-methods} describes the implementation and validation methods.

\subsubsection{Reading actual carrier templates}
The recursive carrier engine represents generics using \texttt{Genericable} objects with \texttt{extendToBe}, \texttt{declaredToBe}, \texttt{hasToBe}, and \texttt{definedToBe} slots. Its calls use constructor-specific dispatch and \texttt{ProjectSession}. The bounded adapter reads actual objects after carrier inference. It assigns reachable generics identifiers keyed by AST identity and object identifier, exports nontrivial value-side lower bounds and declaration/context/structure upper bounds, and preserves symbolic references between exported variables.

The adapter supports selected primitives, first-order function objects, dictionaries, and records with an open base and immutable non-default fields. It rejects unresolved returns, constructor cycles, unknown or mutable record fields, unsupported receiver and higher-order constructors, carrier arrays, and carrier Options. Carrier arrays expose mutation; carrier Option matching is not the kernel's closed selector judgment. Higher-order ground \emph{arrow types} are in the kernel syntax, but not every carrier \texttt{HigherOrderFunction} object can be translated.

Candidate domains contain ground subterms of the actual type and exported bounds, followed by fixed primitive candidates. It limits traversal depth to 32, reachable generic variables to 12, candidate types to 128, and checks to two million. Unsupported conversion is a separate outcome. \textsc{NoSolutionInDomain} describes only that candidate pool.

The adapter is invoked explicitly after inference. Its output certifies the exported template, subject to Table~\ref{tab:claim-correspondence}'s source-generation boundary. Section~\ref{sec:source-residual} instead constructs an export directly from the stated source grammar.

%% file: sections/support/checked-execution.tex
\subsection{Prechecked Evidence and Runtime Validation}\label{sec:checked-execution}

\texttt{CheckedRecordProgram} executes Outline\(_R\) with immutable integer, boolean, string and record actuals. Its closed-program serializer requires field identifiers matching \texttt{[A-Za-z\_][A-Za-z0-9\_]*}; the lower-level source/value API accepts arbitrary string record labels. It serializes the accepted function and actuals into a closed program, then owns the final AST shared by \texttt{SourceResidual.compile} and \texttt{OutlineInterpreter}. Its phases and failure interface are:

\noindent\begin{minipage}{\linewidth}
\begin{verbatim}
prepare(function, actuals, candidates, budget):
    parse the closed pure program; export E from that AST
    capture actual values; project T_rho(E) with Unit
attempt(prepared, submittedTrace):
    check C; otherwise EvidenceRejected (before execution)
    build explanations privately; require carrier inference
    run the owned AST; catch InterpreterException
    evaluate the pure body; catch ReferenceException
    compare results; otherwise ResultMismatch
    check the observed value's coverage; otherwise CoverageFailure
    return Success(value, type, explanations)
\end{verbatim}
\end{minipage}
B establishes formal value coverage before evaluation. The Java execution phase then compares the observed result with pure evaluation and verifies its coverage. Each failure returns the named status; Success publishes the value and explanation together. The pure fragment uses immutable values and has no external effects. Unsupported or overflowing Java \texttt{Long}-to-\texttt{Integer} conversion returns CoverageFailure. The convenience \texttt{execute} method throws on failure; the experimental interface retains structured statuses.

\paragraph{Downstream task: an explanation of a successful field obligation}
A client asks which source occurrence required a record component and which specialized comparison discharged it. \texttt{SourceEvidence.explain} returns an ordered list
\[
 (j,o,\mathit{expression}_o,\mathit{location}_o,p,L,U),
\]
joining each submitted leaf with its original source root \emph{after} C succeeds. Each entry describes a component of the specialized subtype comparison. An empty path denotes a whole-root shortcut; a rootless program has no entries.

The consumer guarantee follows directly from Theorem~\ref{thm:residual-origins}(3): every published entry has the original occurrence, expression and location, and the path and endpoints of that occurrence's specialized decomposition. Conversely, every canonical leaf is represented with its multiplicity. Projecting the equal ordered traces and looking up immutable roots proves both directions.

The wrong-path and wrong-endpoint examples in Section~\ref{sec:source-residual} become incorrect component descriptions through an unchecked consumer, despite valid original bounds. C blocks them. Rebuilding from the trusted export also gives correct current descriptions; the receiving contract below additionally asks for a verdict on the saved assertion.

\paragraph{Auditing a saved assertion versus refreshing a display.}\label{sec:saved-claim}
Our implemented receiving task takes an immutable saved claim \((id,\mathit{trace})\) and trusted context \((E,\rho,\eta,j)\) with captured values. It must return a verdict \emph{about that claim}, linked to its ID, while preserving the submitted file. This models reviewing a persisted specialization report or evidence handed from one module to another.

\texttt{SourceEvidence.audit} applies B and C and returns Verified, Rejected, or InvalidCertificate; only Verified carries the submitted claim's checked explanations. \texttt{SourceEvidence.regenerate} applies B and reconstructs from the original export, returning Regenerated with current explanations or InvalidCertificate. Regenerated deliberately says nothing about whether the old assertion was true. \texttt{ClaimTransport} checks that the output does not name the saved input file, including through a hard or symbolic link, then reads the claim, invokes the selected API and writes a separate receipt. Aliased output paths raise a transport error before writing. The reconstruction-and-comparison route invokes B through \texttt{regenerate}, reconstructs true leaves from its complete explanations and compares them with the saved payload in order. It returns the same Verified/Rejected/InvalidCertificate contract as C. The comparison, including truth flags and multiplicity, is the distinguishing operation; regeneration followed by that operation is an implementation of C.

The transformation obligation precedes this final equality test. Source export must distinguish equal constraints generated at different occurrences. Prefix capture must retain roots that become true, and copying must rename sessions without changing occurrence identity. Decomposition then expands each retained root into its ordered paths and endpoints, including an explicit successful leaf for an empty required row. Theorem~\ref{thm:residual-origins} connects these stages, so rebuilding from the original export and decomposing the correctly formed residual give the same occurrence-indexed report.

\paragraph{The report-reuse relation.}\label{sec:report-relation}
For a completed context $X=(E,\rho,\eta,j)$, put $\theta=\rho\uplus\eta$ and let $\mathcal L(X)$ be the ordered canonical trace reconstructed from all roots of $E$ under $\theta$. Suppose a producer context $X_0$ issued the saved trace $\mathcal L(X_0)$. The receiving workflow first checks the mapper syntax and report key and establishes the occurrence correspondence. Let $\nu_{0\to1}$ rename only its session prefix. For an original-valid current certificate, the reuse condition is
\begin{equation}\label{eq:report-reuse}
 \operatorname{Reuse}(X_0,X_1)
 \quad\Longleftrightarrow\quad
 \nu_{0\to1}\mathcal L(X_0)=\mathcal L(X_1).
\end{equation}
Equality compares every root occurrence, path, endpoint and truth value in order, including multiplicity. This is precisely C's payload comparison after import. Theorem~\ref{thm:residual-origins}(1--2) makes $\mathcal L(X)$ independent of how a fixed total substitution was partitioned into capture stages. Equation~\eqref{eq:report-reuse} then compares the resulting reports across successful specializations: changes to inputs or candidate order matter through their effect on this trace.

\paragraph{Receiving with a retained footprint.}\label{sec:frame-receiver-contract}
The file receiver prepares one saved claim by checking it against the old certificate and retaining its bytes, verified receipt, original export, namespace, old ground root comparisons and record-valued cuts. A new request reads the submitted claim and matches those bytes, export identity and namespace. If they differ, it decodes the submission and uses reconstruction plus comparison. Otherwise it checks B on the current certificate. An invalid certificate yields the usual InvalidCertificate receipt. For a valid one, it materializes the current root comparisons and checks that the old and new pairs are in the ground record fragment, with unchanged roots and right endpoints and with width-extended left endpoints. A failed guard invokes reconstruction plus comparison.

If the guard succeeds, the receiver compares the subtrees at the retained record-valued cuts. By Corollary~\ref{cor:record-cuts}, this decides equality to the prepared claim. In the equal case it can emit the retained verified explanations because the same export and namespace also fix their expression and location metadata. In the unequal case it emits Rejected with an empty explanation list. The file adapter writes a separate receipt using the same encoding as reconstruction. This establishes equality of the returned receipt for each decoded claim under the same trusted context and stable file bindings. 

%% file: sections/support/consumer-contract.tex
\subsection{Approval of a Returned-Field Assertion}\label{sec:consumer-contract}

A consumer chooses which fact it approves. We specify a consumer of a body
\(e=h.k\) in Outline\(_R\) and a closed wire schema \(S\). A schema is a
record type built from signed 32-bit integers, booleans, strings and nested
records. A value satisfies \(S\) when it has precisely the declared keys
and recursively satisfies their declared shapes. We write this as
\(\mathsf{shape}(v)=S\), where \(\mathsf{shape}\) records all keys and
uses the exact primitive kind. For the wire task, serialization preserves
keys and primitive kinds recursively. The implemented encoder rejects
unpaired surrogates and escapes Unicode pairs, quotes, backslashes and
control characters. Delivery encodes the value of the checked body under
the same immutable captures. A rejected encoding yields
\texttt{ENCODING\_FAILED} with no payload.

The review store binds an approval to the immutable export \(E\), its
session namespace \(j\), the output schema \(S\), and one true equality
leaf at the outermost read \(o\) of \(h.k\). The leaf may compare \(S\)
with \(S\) at field path \(k\), or compare \(\{k:S\}\) with \(\{k:S\}\)
at the root \(\epsilon\). This admits \texttt{r -> r.box} when the
host is exactly the singleton record required by the read. Preparation
checks B, canonical leaf membership and exact output shape. An approval
of a field leaf stays tied to that leaf when a later comparison stops
at the root. Current delivery checks B, the saved leaf and exact shape.

The publisher states what a review authorizes. A consumer of one named equality can use \emph{Predicate}. A consumer who must cite that equality at a particular step in the derivation can use \emph{Leaf}. A publisher who attaches a review ID to the complete displayed derivation uses \emph{Whole}. That ID covers the ordered entries at every source occurrence, including multiplicity, endpoints and paths. If one inner equality becomes several field comparisons, the output leaf can survive while the displayed derivation requires a new review.

For whole-report approval, preparation also checks that the saved report
is canonical for the old certificate and binds its ID and complete ordered
contents. A delivery request must present that same report. A replacement
under the same ID is rejected before checking whether the saved report
still matches the current certificate.

For process restart, the trusted store saves a versioned structural
encoding of the frozen export and namespace, the schema, the ordered
report and the approved leaf's index. It records parameter order,
ordered candidate lists, source occurrences and offsets, type trees
and the frozen expression. Restoration compares this definition encoding
and schema with the current context and reconstructs the approval on
the current export object. The producer's preparation supplies the old
checks and the receiver checks the current certificate. Exact encoding
equality gives the same frozen definition and source identities, so the
leaf and shape tests apply as before. For whole-report delivery the
receiver can apply the capture-domain rule below when it controls the catalogue.
The review store and the current catalogue are trusted inputs.

\paragraph{Choosing the approved object.}\label{sec:approval-contract-choice}
A direct assertion checker is an alternative to report membership. Bind a reviewed equality to \((E,j,o,p,A,A)\), where \(p\) is a structural field path from original root \(o\), and \(A\) is its approved endpoint. For a current ground comparison \((a_X,u_X)\), define
\[
 \mathsf{Predicate}(X)\quad\Longleftrightarrow\quad
 a_X|_p\text{ and }u_X|_p\text{ are defined and both equal }A.
\]
This predicate remains relative to the selected certificate. Its path is evaluated in the ground types even if the comparator stops at an ancestor. A gate can check B, this predicate, exact output shape and encoding, without requiring any report leaf at \(p\). For the same bound equality, leaf membership additionally requires every proper prefix of \(p\) to expand; whole-report identity additionally preserves all other leaves. Thus the evidence tests imply
\[
 \mathsf{Whole}(X)\ \Longrightarrow\ \mathsf{Leaf}(X)
 \ \Longrightarrow\ \mathsf{Predicate}(X),
\]
with strict implications on arbitrary B-valid edits. For \texttt{r -> r.box}, removing a spare input field preserves the equality at \(box\) but moves the canonical cut to the root. For \texttt{r -> \{box=\{k=r.k\},spare=r\}.box}, adding an unread input field preserves the outer leaf but changes the inner report. Section~\ref{sec:consumer-evaluation} checks both separations on these two inputs. Under width extension with fixed read bindings in Theorem~\ref{thm:report-frame}'s fragment, the old prefixes remain expanding, so the predicate and leaf tests for this bound equality coincide. Their difference matters outside those premises.

For the inventory in Section~\ref{sec:running-example}, changing the string at \texttt{debug} can preserve every type comparison. Approving that string requires a value-level check. For a Whole review, direct reconstruction followed by equality with the saved report implements the receiving decision for any supported B-valid input. On an eligible width edit, Corollary~\ref{cor:occurrence-impact} also supplies the old comparison and the insertion that invalidated it. The smaller set \(F_E\) answers the whole-report question if the consumer only needs the verdict. Output shape and successful encoding remain separate conditions for delivery.

\paragraph{Approval of a published report.}\label{sec:approval-workflow}
In our release prototype the engineer approving an inventory mapper records the report ID with the frozen export, namespace and output schema. A downstream auditor can use the ID to retrieve the approved explanation and cite a particular source occurrence, path and pair of compared types. In Express~4.17.3 the dependency entry states equality with the complete 30-member row. In 4.18.0 the mapper still returns the same range, but its dependency entry expands into 30 separate comparisons. Republishing that explanation under the original approval would silently change what the auditor's citation establishes. The receiver therefore binds the ID to the ordered contents, carries it forward on equality, and requires a new approval to publish changed contents as reviewed. The auditor may still cite the historical report for the earlier release. The same-ID replacement control in Section~\ref{sec:consumer-evaluation} checks this binding.

The inventory owner can instead approve just the returned-field assertion under Predicate or its particular occurrence and path under Leaf. Those contracts suffice when the downstream reader cites only that assertion. A service accepting any valid witness can also retain an old read assignment still admissible in its catalogue (Section~\ref{sec:approval-replay}). A service publishing a fresh solver result checks the report of the assignment actually selected.

\paragraph{Exact shape from the source.}
Let \(\rho\) record the exact types of the captured immutable values.
The shape expression \(F(e)\) is a tree of record constructors with
leaves that are either primitive kinds or \((i,q)\), a field path \(q\)
from captured parameter \(i\). A parameter produces \((i,\epsilon)\).
A literal produces its primitive kind. A record constructs a shape
expression from its field expressions. A field read projects from this
expression: projection on a constructed record selects its named
child, and projection on \((i,q)\) appends the field to \(q\).
Resolve each \((i,q)\) by reading \(\rho(p_i)|_q\), and call the result
\(X_\rho(e)\). Equivalently, an eager interpreter computes a shape at
each source node, looks up a field in its exact host shape, and uses
that result at a read. The implementation uses this eager form.
Selected read-variable bindings do not enter the computation.

\begin{gtheorem}[Single-assertion approval and delivery]\label{thm:consumer-approval}
Fix a prepared approval as above and a current B-valid context for
the same \(E,j\). Let \(p\) be the decomposition path of the approved
leaf, and write \((a',u')\) for the current ground comparison at root \(o\).
\begin{enumerate}
\item The saved leaf is present exactly when every proper prefix of \(p\)
still expands to the next required record field, and the two current
endpoints at \(p\) equal the two approved endpoints.
\item The current result \(v\) of the pure body has
\(\mathsf{shape}(v)=X_{\rho'}(e)\). Whenever encoding succeeds, the wire
payload satisfies \(S\) exactly when \(X_{\rho'}(e)=S\).
\item The approval survives exactly when the leaf test in (1) succeeds
and \(X_{\rho'}(e)=S\).
If the current captures are width extensions with fixed read bindings in
the fragment of Theorem~\ref{thm:report-frame}, the path test reduces to
\(a'|_p=a|_p\), where \(a\) is the old lower endpoint. In the shape test
it suffices to compare \(\rho'(p_i)|_q=\rho(p_i)|_q\) at the capture leaves
\((i,q)\) of \(F(e)\).
\end{enumerate}
\end{gtheorem}

\input{sections/support/consumer-proof.tex}
\paragraph{Consequences for the receiver.}
On an arbitrary valid edit, a changed endpoint, missing edge, or new
cut above the approved path can invalidate membership of the approved leaf. An edit can
also retain the approved leaf while adding a field to a captured subtree
that reaches the output. Changes to unapproved leaves preserve approval
when the approved leaf and exact output shape remain unchanged. These
distinctions permit a receiver to replay
only the approved path in the comparison, while checking the actual
shape independently of inference. B still checks all original bounds.

\subsubsection{Approvals across catalogue changes}
Write \(E=(I,D)\), where \(I\) contains the ordered parameters, the
original roots and their source metadata, the symbolic result type and
the frozen body. The catalogue \(D\) supplies a list for each variable.
Here \(D\) is the catalogue saved in the approval record and \(D'\)
is the receiver's current catalogue. All restoration policies require
the same namespace, schema, \(I\) and variable keys. Exact restoration
also requires \(D=D'\). The append policy keeps each saved capture
list as a prefix and each ordered read list unchanged. It preserves
admission and list positions for every capture type registered at
approval time, so a reviewer can require the current catalogue to
honor that inventory. The prefix is relative to the saved approval. For example, a release gate can require that every previously admitted producer version remain admissible during a staged rollout, with candidate positions preserved for tools that consume ordered capture lists. Append supplies that catalogue-level commitment. A gate that authorizes each request against the currently active versions can use Current. Both gates still verify the bound report and the exact output shape for each delivery.
The current catalogue policy checks each request against
the current domains and rechecks its bound report and output.
After capture, projection searches only the ordered read domains.

\begin{gtheorem}[Approval under catalogue changes]\label{thm:capture-migration}
Fix a prepared B-valid whole-report approval and a current context
with the same namespace, schema, \(I\) and variable keys.
Let \(\rho'\) be a completed capture admitted by \(D'\) that
width-extends \(\rho\). Work in the ground fragment of
Theorem~\ref{thm:report-frame}.
\begin{enumerate}
\item An old feasible read assignment whose entries still belong
to their current domains yields a B-valid certificate with the
result instantiated under \(\rho'\). The append policy guarantees
this domain condition for every old feasible read assignment.
\item For any current B-valid certificate, the bound report is
canonical exactly when the read bindings and the old protected
left subtrees agree. Once the chosen policy restores the approval,
delivery succeeds exactly when the submission matches the bound
report, that report is canonical, the exact output shape is \(S\),
and encoding succeeds.
\end{enumerate}
\end{gtheorem}

\begin{proof}
The common \(I\) fixes every symbolic root and its occurrence.
Fix any old feasible read assignment. Its values still belong to
their current domains, and every new capture has passed its domain
check. Each ground right endpoint is fixed; each lower endpoint
width-extends by induction on the parameter and read references,
constants and record constructors in the root. Ground width
preservation gives all original inequalities, and the new return is
instantiated with the same assignment and the new captures.
For the second claim, let \(\widehat D(v)\) list the union of the
candidates in \(D(v)\) and \(D'(v)\) for each variable \(v\).
Both certificates remain B-valid in the common export
\((I,\widehat D)\): their assignments are admitted and their
original bounds and instantiated returns are unchanged. Their reports
depend only on the shared roots, namespace and fixed assignments.
Apply Theorem~\ref{thm:report-frame} to these two certificates in the
common export. The original submission check and
Theorem~\ref{thm:consumer-approval}(2) give the remaining delivery tests.
\end{proof}

A build service that publishes the current solver's certificate can
use the approved report to explain the selected type at each source
read. It obtains the certificate by running projection on the current
capture and catalogue, then checks the report against that result.
B certifies admissibility of the supplied assignment. A delivery
consumer that accepts any valid witness can check the old read
assignment with the new captures. Section~\ref{sec:approval-replay}
includes a width edit for which the old witness remains valid but
fresh projection selects a different binding. For comparisons outside
the cut criterion, the receiver reconstructs the report.

%% file: sections/support/consumer-proof.tex
\subsubsection{Proof of the single-assertion delivery criterion}\label{proof:consumer-approval}

Fix the trusted export, namespace and prepared approval in
Theorem~\ref{thm:consumer-approval}. The same source occurrence
therefore determines the compared root and the returned field.
The prepared tuple is an actual leaf of the old report.

\paragraph{Leaf preservation}
A record path in the canonical decomposition
can be traversed at a pair of records exactly when the pair is unequal,
the required row is nonempty, and the next required label occurs on
both sides. A successful comparison cannot emit a missing-field leaf.
Induction on the strict prefixes of \(p\) establishes that decomposition
visits that path precisely when each of the stated expansion tests
succeeds. At the terminal pair, the two approved endpoints coincide.
Matching both endpoints makes equality the stopping rule, with
the old path and true flag. Conversely, occurrence of the approved
leaf entails a visit to \(p\) and equality of both reported endpoints.
For root equality, \(p=\epsilon\) and the prefix condition is vacuous.
The required endpoint \(\{k:S\}\) makes this an equality about the
singleton host of the approved read.
No other root or path can supply the tuple because occurrences and
encoded field paths are injective. This proves (1). Under width
extension with fixed upper endpoint, Theorem~\ref{thm:report-frame}'s
proper-record argument preserves every old expanding prefix. The
upper subtree at \(p\) is fixed. The terminal lower subtree is thus
the only remaining comparison test.

\paragraph{Shape adequacy}
Use the exact captured shape environment,
which B checks against the immutable values. Induct on the source
body. Variables and literals have exactly the shape assigned to
them. Record construction preserves the exact key set and has the
child shapes by induction. For a field read the induction hypothesis
gives the exact host shape. B and Theorem~\ref{thm:residual-origins}
ensure that evaluation of the read is defined. The required child's
shape is therefore exactly the corresponding field in that host.
The same induction establishes agreement with \(F\): selection from
a constructed record picks its child expression, and selection from
a captured subtree appends one field before resolution. These are
all possible record-producing forms of \(F\). For the wire grammar,
decoding a successful encoding preserves primitive kind and all
record keys by induction on values. Thus the serialized value has
the derived shape and satisfies the closed schema exactly when the
shape equals \(S\). This proves (2).

\paragraph{Approval and edit dependencies}
Preparation has established
\(X_\rho(e)=S\). The policy requires both current leaf membership
and current exact shape \(S\), so (1) and (2) give the equivalence in
(3). The normalized shape expression is fixed by \(e\). Resolving
it with either capture map changes only the capture leaves.
Equality of the resolved trees is equivalent to equality at those
leaves. This follows by induction on \(F\) because record
construction with fixed, distinct labels is injective in its child
types, and constant leaves do not depend on captures. The proof
also covers an expression with no capture leaves, whose exact
shape is constant. Width extensions keep all the old paths
defined. Therefore the single comparison subtree and the capture
subtrees named by \(F(e)\) give exactly the width-edit tests for
this approval. \hfill\(\square\)

%% file: sections/02b-route-evaluation.tex
\section{Evaluation of Report Preservation and Delivery}\label{sec:route-evaluation}

We compare the source-path predictions with report reconstruction, apply the approval contracts to delivery requests, and measure the costs at each stage.
\begin{enumerate}
\item \emph{Affected comparisons.} We predict the affected source occurrences and compare them with freshly reconstructed reports. The nested-record control tests when a whole-report verdict can conceal different sets of affected comparisons (Section~\ref{sec:footprint-cost}). The PodSpec and API Extractor histories test predictions on published changes (Table~\ref{tab:pod-schema} and Section~\ref{sec:rush-review}). Section~\ref{sec:mixed-evaluation} tests deletions and read reselection.
\item \emph{Delivery under saved approvals.} An inventory mapper separates changes to a returned dependency range from changes to its type report (Table~\ref{tab:express-releases}). We then vary the approved object and catalogue policy while holding the request fixed (Tables~\ref{tab:consumer-outcomes} and~\ref{tab:approval-migration}). In publication-order replay, the receiver records updates to its comparison baseline separately from simulated approvals of changed reports (Table~\ref{tab:prefix-online}).
\item \emph{Preparation and receiving costs.} Table~\ref{tab:footprint-four-times} measures preparation and lookup with fixed reads and validated width premises. Table~\ref{tab:approval-migration-cost} starts with a current certificate. Table~\ref{tab:approval-cycle} includes fresh projection, file operations and fallback.
\end{enumerate}

The source suite includes 21 written programs and a separate 107-case boundary suite. The width survey crosses three mappers, three edit locations and two candidate orders with six same-shape order values, giving 108 pairs. The catalogue comparison crosses 12 scenarios with those six values and adds one 32-read case. Its 84 timing blocks come from four cases, seven blocks and three JVM forks. The 189,217 bounded ground comparisons allow the right endpoint to vary. The separate source-footprint study counts insertion decisions over collections of roots that share inputs.

\input{sections/support/express-releases.tex}

\subsection{Source-footprint synthesis}\label{sec:source-footprint-evaluation}
An independent Python implementation compiles Corollary~\ref{cor:source-footprint}'s capture footprint and compares its verdict with canonical report reconstruction. Eight configurations of seven source programs exercise repeated reads, shared parameters in constructed records, ancestor subsumption, separate parameters, fixed read hosts, rootless exports and primitive/top cuts. For 36 single insertions and 66 two-step edits, the checker derives insertion sites from the initial and final captures and tests whether a protected path is a prefix of a site. This prediction and direct subtree comparison both agree with reconstruction, including when the second field is inserted inside a newly added record. Seven separating edits change one retained test while preserving the others. Exhausting the 48 candidate conjunctions over these controls' old capture-record paths finds 10 equivalent conjunctions; each normalizes to the synthesized footprint.

A bounded check uses eight symbolic lower expressions over 133 paired parameter assignments from the 133-type universe. Each successful collection has two roots sharing an expression and a third root depending on the other parameter. Across 12,969 such collections, reconstruction preserves the report in 19,890 insertion cases and changes it in 61,196. The input-path checks agree in all 81,086 cases. A further 18,123 insertions each change one retained test while leaving the other retained tests intact. A separate cross-language control extends the field read by \texttt{r -> r.k}. Its record has labels U+10000 and U+E000, whose UTF-16 ordering differs from code-point ordering. We insert a third field and freshly project both captures. At each capture, the selected type equals the record at $k$. Java and the independent Python reconstruction each stop at the single path \texttt{/field:k} and agree on the ordered fields of both endpoints. The new endpoint includes the inserted field.

\input{sections/support/footprint-cost.tex}
\input{sections/support/frame-type-evaluation.tex}

\input{sections/support/manifest-impact.tex}

\input{sections/support/release-review.tex}

\input{sections/support/mixed-edits.tex}

\input{sections/support/rush-reports.tex}

\subsection{Report reuse in delivery workflows}\label{sec:reuse-evaluation}
The following studies use checked reports to authorize delivery. They distinguish the evidence that survives an edit from the cost of checking a request. Supplementary material S1 gives the source-export controls, individual operation costs and distributions across JVM runs.

\input{sections/support/source-reuse.tex}

\subsection{Evaluation Scope}\label{sec:evaluation-limitations}

The inventory study follows one read, \texttt{m.dependencies.debug}, across Express, Koa and Mocha. All three packages declare this dependency. Consecutive versions share a release history, so a transition is the unit of observation within each series. Table~\ref{tab:project-series} compares policies on adjacent pairs. Table~\ref{tab:impact-stream} follows a saved report until its comparison reference is renewed. Section~\ref{sec:prefix-review} admits previously observed types in publication order and records simulated approvals. The Chalk study holds an earlier dependency row fixed across a deletion. The PodSpec study covers every adjacent release pair in its stated interval and follows two nested security settings. For both studies, releases supply the edits and our mappers define the comparisons. The API Extractor study follows two declarations across two consecutive release windows. The upstream project reviews public interfaces. Our mapper fixes the original read types and compares the evidence for those two occurrences.

The cross-process study pairs existing order records with designed catalogue changes. The seven relevant commits in the native-example history concern comments, an effectful handler and a source-preserving rename. The controlled source suite uses written programs and generated graphs. In the separate carrier study, port adaptations and label mappings use known benchmark answers, and several benchmark facts can share a binding.

Measurements come from one arm64 macOS machine. The controlled workloads isolate preparation, projection, comparison and delivery. The continuous replay measures local receipt on the three published release series, using its fixed catalogues and an in-memory review store. Each timing table specifies the stages included in its totals.

\subsection{Development and Validation Methods}\label{sec:development-methods}

Java executes the carrier inference and receiving workflows. Independent Python implementations reconstruct the exported constraints, enumerate finite assignments and verify receipts. Section~\ref{sec:footprint-cost} uses Python to compile and time the source-path checks. Supplementary material~S1 describes the broader carrier and its benchmark scoring protocols.

%% file: sections/support/express-releases.tex
\subsection{Replaying approvals across Express releases}\label{sec:express-releases}

To measure reuse under successive dependency changes, we collected the manifest from every stable Express package on npm from 4.0.0 through 4.21.2, in numeric version order \citep{express-npm-releases}. This gives 91 published manifests and 90 consecutive pairs. We retained the package name and the complete \texttt{dependencies} object. Each entry names a dependency and gives its requested version range as a string \citep{npm-package-dependencies}. A change to a range can change the output value while preserving its type. Adding a dependency introduces a field into the captured record. We applied two approval policies to a single inventory mapper over these published manifests.

The mapper reads \texttt{m.dependencies.debug} and returns its range in a one-field record. Its executable body is \texttt{\{box=\{range=m.dependencies.debug\},spare=true\}.box}. Consider a receiver that imports the range cell into a dependency inventory. Under Leaf, review covers the outer \texttt{box} comparison with both endpoints equal to \(\{range:String\}\). Under Whole, review also covers the comparison at \texttt{m.dependencies}. In the old release this read selects the complete dependency row. Consequently the old report contains an equality at the path \texttt{/field:dependencies}. Adding a dependency changes this equality even if the inventory still receives exactly the same bytes.

For each pair we prepared an approval on the first release and ran the Java projector afresh on the second. Both projections use the same export, namespace and ordered candidate list. The list starts with \(String\), the old and new dependency rows in that order, then the output schema and the two capture types. We checked the original bounds and the current canonical report on both sides. For each approval the receiver then checked the current certificate, the saved evidence and the exact output shape before issuing a payload. Python independently reconstructed both reports. For every width-related pair it also compiled the old cuts to input paths and checked the verdict against the Java replay.

\input{sections/support/express-table.tex}

Table~\ref{tab:express-releases} shows two directions in which the evidence and the delivered value can vary independently. Each entry counts release pairs. Equal payload compares the encoded range cells, and Leaf and Whole count deliveries under the corresponding saved approval. Among the 78 pairs with the same dependency names, ten changed the requested range for \texttt{debug}. All 78 kept the complete report. In all ten strict additions the solver retained the old read types and the returned-field approval survived, but the complete report changed. Six of those ten delivered the same range bytes under Leaf. For example, 4.18.0 added \texttt{http-errors} to the dependencies of 4.17.3 and kept \texttt{debug} at \texttt{2.6.9}. The old dependency-row equality was the part of the review that expired. The compiled footprint locates this change at \texttt{m.dependencies}. The two pairs with a removal were decided by canonical reconstruction; fresh projection changed a read binding in both.

Across this release series the saved Whole approval covered 78 deliveries and the saved Leaf approval covered 90. The difference comes from the scope of review in this mapper. Leaf covers the returned range cell. Whole also covers the complete dependency row chosen on the old release.

\paragraph{Sensitivity to the selected dependency type.}
We applied the same mapper to every stable Koa release from 2.0.0 through 2.15.4 and every stable Mocha release from 10.0.0 through 10.8.2, ordered numerically within each range \citep{koa-mocha-registry}. We chose these additional series because the same field is present throughout, allowing us to change the project while holding the read expression fixed. We verified the published package tarballs and extracted their declared dependencies. For each pair the full-row catalogue includes the observed old and new dependency rows. The one-field policy also admits \(\{debug:String\}\) and places it ahead of the full rows. Both catalogues satisfy the read of \texttt{debug}. Each projection starts afresh. Java checks and delivers the approved output cell and separately checks Whole; Python reconstructs the two reports and, when applicable, independently checks the footprint.

\input{sections/support/project-series-table.tex}

In Table~\ref{tab:project-series}, Width includes unchanged key sets and additions, while Strict counts additions alone. Whole counts pairs that preserved the complete report after fresh projection. Among strict additions, the full-row-first catalogue preserved 0/11 reports and the one-field-first catalogue preserved 11/11. Seven pairs removed dependencies and fell outside the width-extension criterion. We checked their reports by reconstruction under both catalogues. Section~\ref{sec:mixed-evaluation} tests the per-occurrence criterion on the same seven pairs. Only those seven pairs changed the full-row selection. The one-field selection held throughout. Mocha had no strict additions. Table~\ref{tab:project-series} shows how often the width premise held for these manifests and how Whole depends on the reviewed binding. With the returned cell as the unit of review, Leaf authorizes every transition under both catalogues.

\paragraph{How the read catalogue selects the reviewed assertion.}
The projector accepts an ordered finite domain for each unknown. Forward checking discards inconsistent candidates; search visits the retained candidates in the supplied order and returns the first complete solution (Algorithm~\ref{alg:projection}). The two read-catalogue policies in Table~\ref{tab:project-series} are experimental inputs. Selecting the complete old row records equality of the dependency names and their field types at that read. Both records can still contain different version-range strings of the same type. A consumer requiring only the capability to read \texttt{debug} can request the one-field candidate. When the old row contains other fields, that choice emits a narrower comparison from the first release onward. If the new order selects a different read type, the right endpoint recovered from the new report changes. In the adjacent-pair experiment, read choices changed on 7 of 138 transitions with the full row first and on none with the field first. The receiver used reconstruction for all seven edits with removals under each policy. Section~\ref{sec:impact-stream} follows a single ordered catalogue through each complete series.

%% file: sections/support/express-table.tex
\begin{table}[tbp]\centering\small
\caption{Approval outcomes across Express releases}\label{tab:express-releases}
\begin{tabular}{@{}lrrrrr@{}}\toprule
Dependency keys & Pairs & Same read types & Equal payload & Leaf & Whole\\\midrule
Unchanged keys & 78 & 78 & 68 & 78 & 78\\
Strict addition & 10 & 10 & 6 & 10 & 0\\
Removal only & 1 & 0 & 1 & 1 & 0\\
Addition and removal & 1 & 0 & 1 & 1 & 0\\
\midrule Total & 90 & 88 & 76 & 90 & 78\\
\bottomrule\end{tabular}\end{table}

%% file: sections/support/project-series-table.tex
\begin{table}[tbp]\centering\small
\caption{Report preservation under two read catalogues}\label{tab:project-series}
\begin{tabular}{@{}lrrrrrr@{}}\toprule
Package & Pairs & Width & Strict & With removal & Whole (full) & Whole (field)\\\midrule
Express & 90 & 88 & 10 & 2 & 78 & 90\\
Koa & 34 & 31 & 1 & 3 & 30 & 34\\
Mocha & 14 & 12 & 0 & 2 & 12 & 14\\
\midrule Total & 138 & 131 & 11 & 7 & 120 & 138\\
\bottomrule\end{tabular}\end{table}

%% file: sections/support/footprint-cost.tex
\subsection{From local obligations to input checks}\label{sec:footprint-cost}
Consider \texttt{r -> \{a=r.k,b=\{box=r\}.box\}} with $r$ of type $\{k:\{v:Int\},spare:Bool\}$. The first read selects $\{v:Int\}$ and protects $r.k$. The second selects the type of $r$. Comparison at its host $\{box=r\}$ stops at whole-record equality, so a check on $r$ covers both obligations.

\paragraph{Four implementations of the same decision}
Rebuild compares the saved report with a freshly reconstructed one. Local keeps one row per obligation, recording its symbolic lower expression, ordered cut paths and old endpoints. Dedup keeps one copy of each distinct row. With read types fixed, identical rows perform the same check on any new capture. Source maps cuts back to input paths and keeps the prefix-minimal paths of Corollary~\ref{cor:source-footprint}. Each occurrence retains its own identity in the saved report.

\input{sections/support/footprint-counts.tex}

\paragraph{Effects of deduplication and ancestor coverage}
In Nested, the lower expressions \(r\) and \(\{box=r\}\) give different prepared rows. Their input paths are \(r.k\) for the first occurrence and \(r.k.v\) for the second. The whole report needs one test, at \(r.k\). Adding \texttt{note} beside \(k\) preserves both reports. Adding a field directly to \(r.k\) fails the global test and changes only the first report. Adding a field directly to \(r.k.v\) fails the same test and changes both. Fresh Java reconstruction confirms the two different affected sets. The distinction is predicted by the local paths in Corollary~\ref{cor:occurrence-impact}.

We checked Equation~\eqref{eq:attribution-from-global} on all six configurations in Table~\ref{tab:footprint-counts}. For each old input we enumerated every subset of its record sites, including the empty subset, adding one fresh Boolean field at each selected site. We grouped these edits by their complete vector of global test outcomes and compared the occurrence reports by reconstruction. Across the 536 subsets, each vector determined one affected set in One, 8 shared, 32 shared and 8 distinct. In each of Ancestor and Nested, one vector occurred with different affected sets. These are batches of insertion sites on six fixed configurations.

Identical shared reads already reduce to one row with Dedup, while eight distinct sibling reads retain eight checks with either method (Table~\ref{tab:footprint-counts}). The strict two-to-one reduction comes from the ancestor implication between \emph{different} obligations. Corollary~\ref{cor:source-footprint} supplies a separating edit for each remaining path under the fresh-label premise.

\paragraph{Preparation and reuse}
For each of the four direct-field mappers we add a field to the capture root (report preserved) or inside the last requested field (report changed). The fifth timed configuration, Ancestor, uses the value edit and the extra sibling described above. All four routes agree on the ten timed requests and the two Nested insertions reported in Table~\ref{tab:footprint-counts}. Table~\ref{tab:footprint-four-times} gives microseconds per operation as the median of three process medians on arm64 macOS with CPython 3.14.4. Each process warms every operation and randomizes operation order in seven measured blocks per mapper. Preparation starts with the old export and bindings and builds each method's saved state. Source also validates its old bounds internally. Per-request times cover the in-memory decision with fixed read bindings and validated width premises. Parsing, projection, catalogue admission, transport and output encoding are separate stages of the receiver.

\input{sections/support/footprint-four-times.tex}

In Ancestor, preparation takes 5.60 microseconds for Source and 3.19 for Dedup; a preserving check takes 0.37 and 1.49 microseconds, respectively. For 32 shared reads the largest reduction already comes from Dedup (17.41 to 1.19 microseconds), with Source at 0.48. These are in-memory decisions with validated premises. Section~\ref{sec:impact-stream} charges the cost of obtaining and validating the edits on complete release streams.

%% file: sections/support/footprint-counts.tex
\begin{table}[tbp]\centering\small
\caption{Subtree test counts before and after compilation}\label{tab:footprint-counts}
\begin{tabular}{@{}lrrrl@{}}\toprule
Configuration & Local & Dedup & Source & Input locations checked by Source \\ \midrule
One & 1 & 1 & 1 & $r.k$ \\
8 shared & 8 & 1 & 1 & $r.k$ \\
32 shared & 32 & 1 & 1 & $r.k$ \\
8 distinct & 8 & 8 & 8 & $r.k_0,\ldots,r.k_7$ \\
Ancestor & 2 & 2 & 1 & $r$ \\
Nested & 2 & 2 & 1 & $r.k$ \\
\midrule
\multicolumn{5}{@{}l@{}}{Adding $r.note$ in Nested preserves the report (all four methods).} \\
\multicolumn{5}{@{}l@{}}{Adding $r.k.note$ in Nested changes the report (all four methods).} \\
\bottomrule\end{tabular}\end{table}

%% file: sections/support/footprint-four-times.tex
\begin{table}[tbp]\centering\small
\caption{Preparation and query times for four reuse methods}\label{tab:footprint-four-times}
\begin{tabular}{@{}llrrrrr@{}}\toprule
Stage & Method & One & 8 shared & 32 shared & 8 distinct & Ancestor \\ \midrule
Prepare once & Rebuild & 1.68 & 12.10 & 46.23 & 13.67 & 2.80 \\
 & Local & 1.50 & 11.10 & 43.57 & 13.25 & 2.74 \\
 & Dedup & 1.85 & 12.64 & 47.93 & 14.59 & 3.19 \\
 & Source & 3.61 & 20.72 & 77.11 & 37.98 & 5.60 \\
\addlinespace[3pt]
Preserving request & Rebuild & 1.73 & 12.67 & 48.34 & 14.19 & 3.06 \\
 & Local & 0.70 & 4.45 & 17.41 & 5.74 & 1.52 \\
 & Dedup & 0.69 & 0.81 & 1.19 & 5.72 & 1.49 \\
 & Source & 0.49 & 0.48 & 0.48 & 3.33 & 0.37 \\
\addlinespace[3pt]
Changing request & Rebuild & 2.31 & 17.08 & 66.48 & 14.96 & 5.37 \\
 & Local & 0.64 & 0.77 & 1.13 & 5.68 & 1.50 \\
 & Dedup & 0.64 & 0.76 & 1.13 & 5.50 & 1.52 \\
 & Source & 0.44 & 0.44 & 0.44 & 3.21 & 0.37 \\
\bottomrule\end{tabular}\end{table}

%% file: sections/support/frame-type-evaluation.tex
\subsection{Cost of Validating the Width-Extension Premise}\label{sec:frame-cost}

The Python measurements in Table~\ref{tab:footprint-four-times} start with a validated width extension. To measure the cost of establishing that premise, a Java type-level prototype evaluates Theorem~\ref{thm:report-frame} on twelve comparisons: six families at widths 32 and 256. A successful old comparison, its correct saved report and materialized ground endpoints are available initially. The upper endpoint and root identity are fixed. We measure six operations separately: first construct the footprint; validate the current fragment and width-extension premises; query the retained footprint; reconstruct the current report and compare it; and each of the last two decisions together with premise validation. A failed premise makes the footprint inapplicable and requires the general audit. The directly timed guarded decisions avoid treating sums of separate medians as whole-task measurements. The table abbreviates validation as G, the footprint query as Q and reconstruction with comparison as R. All times are in microseconds.

The prototype uses the production immutable type representation and equality operation. Its reconstruction agrees with \texttt{SourceResidual.trace} on every measured input; Python independently checks all endpoints, footprints and both reporting policies. Current lower trees are rebuilt before timing, so equality cannot rely on sharing with the old endpoints. The footprint retains paths and references to old endpoint subtrees; premise validation also retains the complete old comparison context. Section~\suppnumber{app:frame-cost} reports serialized storage as compact UTF-8 JSON including those subtrees and per-invocation thread allocation from \texttt{ThreadMXBean}, distinct from retained heap. Parsing, search, type construction, original certificate checking, transport and output I/O are common surrounding tasks and are outside these type-level operations.

Three sequential JVM forks warm each operation 1,000 times and measure seven shuffled blocks of 100 invocations per case. All 252 rows complete. Table~\ref{tab:frame-cost} reports medians of fork medians at width 256; Section~\suppnumber{app:frame-cost} includes both widths, allocation, storage and fork ranges. Tables~\suppnumber{tab:frame-cost-all} and~\suppnumber{tab:frame-ranges} use the same operation names and timing units. Table~\suppnumber{tab:frame-allocation} gives serialized footprint size, including protected subtrees, and thread allocation per invocation in KiB.

\input{sections/support/frame-cost-main.tex}
\input{sections/support/frame-cost-reading.tex}

%% file: sections/support/frame-cost-main.tex
\begin{table}[t]
\centering\small
\caption{Footprint checking costs at width 256}\label{tab:frame-cost}
\begin{tabular}{@{}lrrrrrr@{}}
\toprule
Family & Build & G & Q & R & G+Q & G+R\\
\midrule
Sparse read & 0.12 & 19.91 & 0.04 & 0.25 & 18.94 & 19.72\\
Many required fields & 15.51 & 20.13 & 11.32 & 53.03 & 35.18 & 74.48\\
Equality cut, intact & 9.50 & 21.27 & 9.45 & 19.24 & 30.16 & 39.40\\
Empty-row cut, intact & 0.32 & 22.24 & 10.50 & 11.66 & 36.33 & 39.20\\
Top cut, intact & 0.27 & 23.21 & 11.07 & 11.47 & 35.05 & 37.08\\
Equality cut, changed & 10.97 & 22.26 & 0.12 & 21.44 & 23.35 & 48.65\\
\bottomrule
\end{tabular}
\end{table}

%% file: sections/support/frame-cost-reading.tex
Direct reconstruction R requires no width-extension guard. At width 256, guarded footprint checking has a lower median than R only for the many-field case, 35.18 versus 53.03 microseconds. Sparse reads cost 18.94 versus 0.25, and intact equality cuts cost 30.16 versus 19.24. First construction and retained context add costs before repeated queries. Table~\ref{tab:frame-cost} also retains the separately measured guarded reconstruction operation.

%% file: sections/support/manifest-impact.tex
\subsection{Tracking a reviewed report through consecutive releases}\label{sec:impact-stream}

A receiver can return a location with its decision. At the start of a review it saves the canonical report, its read bindings, its captured types and the prefix-minimal paths \(F\) of Corollary~\ref{cor:source-footprint}. On receipt it freshly projects the version and checks B. It compares the read bindings and traverses the saved and current captures to verify width extension. The same traversal records each old record path \(q\) whose immediate field set gained a label. For an eligible request it returns all pairs \((p,q)\) on the same capture with \(p\in F\) a prefix of \(q\). An empty list preserves the report and carries forward any approval already attached to it. A nonempty list identifies both the protected subtree and the old record where a field was inserted. An ineligible request goes to canonical reconstruction. The guard timing includes this traversal of the two captures.

\paragraph{A persistent review.}
We stream the same Express, Koa and Mocha manifests as in Table~\ref{tab:project-series}, in version order. In that table each adjacent pair has its own approval and candidate list. Here the export and all candidate domains stay fixed throughout each series. The dependency-read domain lists the distinct dependency-row types observed in the series in order of first appearance (Historical row). The second policy places \(\{debug:String\}\) first (One field). The two other reads have singleton domains for the string and the output cell. The capture domain contains the observed manifest types. This retrospective catalogue isolates the effect of a stable selection policy across releases.

Source keeps the saved state after an eligible preservation. On a hit it withholds delivery and uses the changed report as the next comparison reference. On fallback it reconstructs the report, checks equality and refreshes its state. Rebuild compares the two reports on every receipt and renews its reference on a change. A changed report ends the old approval under Section~\ref{sec:consumer-contract}. The replay automatically installs the new reference and assumes approval of that report for subsequent delivery decisions. Both routes check the source and value shapes and encode the range on delivery. Their decisions and payloads agree throughout. Independent Python reconstruction checks each verdict, reference version, protected path and trigger pair.

In Table~\ref{tab:impact-stream}, Clear counts reports preserved by the path test and Longest is the longest such run on one footprint. Hit counts requests with an insertion at or below a protected path. Fallback counts ineligible requests sent to reconstruction. Same includes all preserved reports, including those recognized on fallback. The lower panel reports microseconds.

\input{sections/support/manifest-impact-table.tex}

\paragraph{What was reused.}
Under Historical row, Express~4.3.0 adds \texttt{proxy-addr} at the path protected by the review of 4.0.0, \texttt{m.dependencies}. The receiver withholds delivery. Version~4.3.1 has the same report as the new 4.3.0 reference, so approval of 4.3.0 would cover 4.3.1. By 4.13.0, the selected historical row is narrower than the actual row, and the comparison already expands to fields. Its empty footprint serves the next 32 versions, including four strict additions. Under One field, the reference from 4.12.0 serves 37 versions, including five strict additions. The seven One field fallbacks preserve the report. In the seven Historical row fallbacks the selected read type changes, which changes the recoverable right endpoint by Theorem~\ref{thm:report-frame}.

Of the 262 eligible requests in Table~\ref{tab:impact-stream}, 203 had an empty footprint, 55 had a nonempty footprint with no insertion sites, and four had both a nonempty footprint and insertion sites. In all four requests with both, the insertion occurred at the protected path itself, \texttt{m.dependencies}. A rule that flags any insertion whenever the footprint is nonempty therefore agrees on all 262 eligible release requests. The Nested mapper of Section~\ref{sec:footprint-cost} distinguishes that rule from the prefix test. Its footprint is \(\{r.k\}\). Adding a sibling of \(k\) at \(r\) preserves the report. Adding a field at \(r.k.v\) changes it and yields the pair \((r.k,r.k.v)\). When fields are added at \(r\), \(r.k\) and \(r.k.v\) together, the receiver returns exactly \((r.k,r.k)\) and \((r.k,r.k.v)\). The Java control checks the sites and pairs against these expectations and the verdict against canonical reconstruction.

\paragraph{Cost per revision.}
Each package and order ran in three fresh JDK~21 processes on arm64 macOS. After two warm-up streams per route, each process ran seven paired blocks with randomized route order. Table~\ref{tab:impact-stream} gives the median of three process medians. A block's initial preparation includes reading the first manifest, projection, B validation, reconstruction and, for Source, footprint synthesis. Receipt total averages the elapsed time over subsequent revisions, including local input-file reading, fresh projection, B validation, decision, delivery on success and renewal when due. The guard column includes read comparison and verified edit extraction; query averages the prefix test over eligible receipts. Each policy uses one in-memory review store. Shared catalogue construction, human review and network transfer are outside the timers. Table~\ref{tab:approval-cycle} separately measures a serialized approval cycle. For Express with historical rows, Source spends about 12 microseconds per receipt verifying reads and width and 0.60 microseconds per eligible query, while its complete receipt takes about 496 microseconds against 461 for Rebuild. For the Express hit at version~4.3.0, the two routes agree on the changed report. Source also returns the protected dependency row and its insertion site.

%% file: sections/support/manifest-impact-table.tex
\begin{table}[tbp]\centering\small
\caption{Report reuse with a fixed catalogue}\label{tab:impact-stream}
\begin{tabular}{@{}llrrrrrrr@{}}\toprule
Package & First candidate & Requests & Clear & Longest & Hit & Fallback & Reselected & Same\\\midrule
Express & Historical row & 90 & 85 & 32 & 3 & 2 & 2 & 85\\
Express & One field & 90 & 88 & 37 & 0 & 2 & 0 & 90\\
Koa & Historical row & 34 & 30 & 14 & 1 & 3 & 3 & 30\\
Koa & One field & 34 & 31 & 14 & 0 & 3 & 0 & 34\\
Mocha & Historical row & 14 & 12 & 11 & 0 & 2 & 2 & 12\\
Mocha & One field & 14 & 12 & 11 & 0 & 2 & 0 & 14\\
\bottomrule\end{tabular}
\par\medskip
\begin{tabular}{@{}llrrrrrrr@{}}\toprule
& & \multicolumn{2}{c}{Initial preparation} & \multicolumn{3}{c}{Source phases} & \multicolumn{2}{c}{Receipt total}\\
Package & First candidate & Rebuild & Source & Project+B & Guard & Query & Rebuild & Source\\\midrule
Express & Historical row & 261.29 & 315.58 & 424.80 & 11.83 & 0.60 & 461.33 & 496.20\\
Express & One field & 382.79 & 442.92 & 438.89 & 8.24 & 0.34 & 492.47 & 492.71\\
Koa & Historical row & 322.38 & 396.50 & 217.40 & 14.99 & 1.06 & 296.66 & 297.05\\
Koa & One field & 594.38 & 441.17 & 334.29 & 15.46 & 0.62 & 449.43 & 431.37\\
Mocha & Historical row & 1,194.71 & 1,074.58 & 648.69 & 20.85 & 0.81 & 863.25 & 802.93\\
Mocha & One field & 1,192.79 & 973.29 & 527.49 & 21.46 & 0.82 & 742.21 & 667.70\\
\bottomrule\end{tabular}
\end{table}

%% file: sections/support/release-review.tex
\subsection{Source comparisons across PodSpec revisions}\label{sec:pod-review}

A revision of a nested schema can add fields at several record paths in the same release. We took the OpenAPI descriptions shipped with all fourteen Kubernetes minor releases from v1.20.0 through v1.33.0, using the first release of each minor version \citep{k8s-pod-schema}. From each description we extracted every property name of \texttt{PodSpec}, following its \texttt{securityContext} reference and the nested \texttt{windowsOptions} reference. Each other property is represented by a Boolean presence marker. The resulting input is a descriptor of published field names. Defaults, property values, requiredness and the internal structure of other referenced types fall outside this view. We supplied the following two-view mapper and its review contract:
\[
\texttt{s -> \{context=s.securityContext, descriptor=\{spec=s\}.spec\}} .
\]
For each adjacent pair we prepared a review on the earlier schema. The first read \(c\) selects the complete old security context. The second read \(d\) selects a descriptor requiring only \texttt{securityContext.windowsOptions} with its complete old row. Both projections use one export admitting the pair's capture types, with the two read domains fixed to these single choices. All thirteen pairs pass B and the width and read guards.

\paragraph{From an insertion site to a source occurrence.}
Corollary~\ref{cor:occurrence-impact} gives two predictions to check on each pair. The global frontier \(F_E\) predicts equality of the complete report. The links \(W_o\) predict the set \(A(I)\) of changed occurrences and identify the local comparison stop, the input path it protects and the insertion site responsible for each change.

Write \(P\) for the PodSpec root, \(C\) for its security context and \(W\) for the Windows options inside \(C\). The old cuts for \(c\) and \(d\) pull back to \(C\) and \(W\), respectively. Dedup keeps both local tests because their expressions and cuts differ. Source uses the single input test at \(C\) for report identity and retains both occurrence links. We checked every predicted set of changed occurrences against a fresh Java report grouped by its original source IDs. A separate Python reconstruction also checked the two local decisions, the Dedup verdict and the source verdict on each pair.

\input{sections/support/pod-schema-table.tex}

Table~\ref{tab:pod-schema} lists the seven transitions with additions by their new version. The other six preserve all three field sets. Three of the seven put fields only at \(P\). All three preserve both reports although a rule that flags any addition whenever the frontier is nonempty rejects them. The addition of \texttt{hostProcess} at \(W\) in v1.22.0 reaches both \(c\) and \(d\) because both old cuts compared an entire record whose subtree has now grown. The v1.32.0 edit adds \texttt{resources} at \(P\) and \texttt{seLinuxChangePolicy} at \(C\). The verified sites are \(I_s=\{P,C\}\). The link for \(c\) ends at \(C\), which covers the insertion at \(C\). The link for \(d\) ends at \(W\), which covers neither site. Thus \(A(I)=\{c\}\), whereas the single insertion at \(W\) in v1.22.0 gave \(A(I)=\{c,d\}\). Both edits hit the same global frontier. A reviewer can therefore reopen the equality for \texttt{s.securityContext} and retain the comparison for \texttt{\{spec=s\}.spec}. These are outcomes for the stated descriptor contract on published schema revisions.

\subsection{Receiving with a catalogue built in publication order}\label{sec:prefix-review}
Table~\ref{tab:impact-stream} fixes its candidate catalogue using the complete series. We repeated the six npm streams in registry publication order with an initially empty collection of observed capture and dependency-row types. Before projecting each release, the receiver admits its current capture and appends any previously unseen dependency row. One field starts with the fixed \texttt{debug} row ahead of observed rows. The other two read domains are singletons. Thus the first request sees only its initial policy and types observed by that release.

The source and its occurrence numbering are fixed. On catalogue growth, the receiver constructs the enlarged export with the same roots, body and return, and validates the saved witness on its old capture in that export. The unchanged roots and old substitution reconstruct the same saved report. Comparing that witness with the fresh projection now uses a common export, as required by Theorem~\ref{thm:report-frame}. On an eligible preservation the receiver keeps its reference. When a changed report ends the previous approval, the replay records a simulated new approval for subsequent requests. A fallback that reconstructs an equal report refreshes the reference while carrying forward the existing approval. In Table~\ref{tab:prefix-online}, Rows gives the initial and final dependency-read domain sizes, Grow counts later expansions, and Project includes initialization. Clear counts requests that retain the current comparison reference, and Longest is the longest run of such requests. Reset counts reference renewals, and Approve counts simulated approvals of changed reports.

\input{sections/support/prefix-online-table.tex}

Across the six streams, 258 requests retained the reference, four eligible hits changed the report, and fourteen requests took the fallback route. Seven fallbacks changed the report. These events produced eighteen reference renewals and eleven simulated new approvals after initialization. Every reference version, read-agreement flag and report decision matched the retrospective run on the same releases. For Express under Historical row, the first projection had one observed row and the final projection had thirteen. A prospective run can thus reproduce the observed decisions with the growing catalogue on these inputs.

We measured the phases in Table~\ref{tab:prefix-online} in three fresh JDK~21 processes per stream on arm64 macOS. Each process performed a separate audit run, two warm-up streams and seven measured streams. Each cell is the median of three process medians of the per-stream mean per invocation of that phase. Times are in microseconds. A dash marks a phase that the stream never invoked. Build covers footprint synthesis and reference construction at initialization and renewal. Guard compares read choices and verifies the width edit while extracting sites. Query performs the prefix test. Fallback reconstructs and compares the report. On an eligible hit the receiver constructs the new reference report directly. Reuse delivery checks the output shapes, evaluates the body and encodes a delivery on an eligible preservation. These phase measurements exclude catalogue growth, its witness check, file reading and human review. Table~\ref{tab:impact-stream} gives the separately measured complete receipt under a fixed catalogue. Source-occurrence explanations in Table~\ref{tab:pod-schema} were checked for accuracy without timing them.

%% file: sections/support/pod-schema-table.tex
\begin{table}[tbp]\centering\small
\caption{Affected source comparisons across Kubernetes PodSpec revisions}\label{tab:pod-schema}
\begin{tabular}{@{}llcll@{}}\toprule
New version & Added properties & Sites & Recheck occurrences & Whole report\\\midrule
1.22.0 & \texttt{hostProcess} & W & c, d & changed\\
1.23.0 & \texttt{os} & P & none & preserved\\
1.25.0 & \texttt{hostUsers} & P & none & preserved\\
1.26.0 & \texttt{resourceClaims}, \texttt{schedulingGates} & P & none & preserved\\
1.30.0 & \texttt{appArmorProfile} & C & c & changed\\
1.31.0 & \texttt{supplementalGroupsPolicy} & C & c & changed\\
1.32.0 & \texttt{resources}, \texttt{seLinuxChangePolicy} & P, C & c & changed\\
\bottomrule\end{tabular}\end{table}

%% file: sections/support/prefix-online-table.tex
\begin{table}[tbp]\centering\small
\caption{Reference renewals and approvals as catalogues grow}\label{tab:prefix-online}
\begin{tabular}{@{}llrrrrrrrr@{}}\toprule
Package & Order & Rows & Grow & Project & Clear & Longest & Fallback & Reset & Approve\\\midrule
Express & Historical & 1--13 & 12 & 91 & 85 & 32 & 2 & 5 & 5\\
Express & One field & 2--14 & 12 & 91 & 88 & 37 & 2 & 2 & 0\\
Koa & Historical & 1--5 & 4 & 35 & 30 & 14 & 3 & 4 & 4\\
Koa & One field & 2--6 & 4 & 35 & 31 & 14 & 3 & 3 & 0\\
Mocha & Historical & 1--3 & 2 & 15 & 12 & 11 & 2 & 2 & 2\\
Mocha & One field & 2--4 & 2 & 15 & 12 & 11 & 2 & 2 & 0\\
\bottomrule\end{tabular}\par\medskip
\begin{tabular}{@{}llrrrrrrr@{}}\toprule
Package & Order & Project+B & Build & Guard & Query & Fallback & Rebuild on hit & Reuse delivery\\\midrule
Express & Historical & 272.00 & 39.83 & 7.49 & 0.60 & 26.08 & 21.69 & 5.05\\
Express & One field & 268.85 & 35.32 & 12.35 & 0.30 & 20.46 & -- & 4.65\\
Koa & Historical & 274.92 & 45.00 & 15.81 & 0.85 & 33.40 & 27.62 & 11.33\\
Koa & One field & 262.47 & 33.92 & 15.68 & 0.70 & 29.69 & -- & 11.19\\
Mocha & Historical & 393.72 & 43.92 & 16.71 & 0.65 & 46.48 & -- & 18.17\\
Mocha & One field & 472.04 & 35.78 & 15.85 & 0.61 & 50.02 & -- & 16.75\\
\bottomrule\end{tabular}\end{table}

%% file: sections/support/mixed-edits.tex
\subsection{Reusing comparisons when dependencies are removed}\label{sec:mixed-evaluation}

Corollary~\ref{cor:surplus} predicts a deletion that leaves the types at every old leaf unchanged and moves the comparison to their parent. We checked it on a published dependency history and then tested the per-occurrence criterion against full reconstruction under two read policies.

\paragraph{A deletion that preserves every old leaf endpoint.}
We screened the npm histories of Debug, Chalk and Cross-spawn for a release returning to an earlier nonempty dependency field set by strict deletion. Chalk supplies a witness \citep{chalk-guard-releases}. We took the dependency names and range strings from the integrity-checked package manifests. Our mapper reads \texttt{m.dependencies}. Its sole read candidate is the record of dependency names with string types in Chalk~1.1.2. The export admits the three tested manifest types as captures. The releases provide the edits and the first release fixes the required row for both transitions.

Chalk~1.1.2 has three dependencies. Version~1.1.3 adds \texttt{has-ansi} and \texttt{strip-ansi}. Java reconstruction replaces one equality at \texttt{/field:dependencies} with three field leaves. Version~2.0.0 removes both extra fields. At the three old leaf paths, the actual and required field types are still \texttt{String}. The new row now equals the required row, so Java emits one equality at their parent. On this second transition the old endpoint tests all pass, and the surplus test alone rejects reuse. On the first transition the old endpoint test already detects the change.

\paragraph{Local decisions after independent projection.}
We reused the 138 adjacent Express, Koa and Mocha release pairs from Table~\ref{tab:project-series}. Under each read catalogue we compared the new criterion against fresh Java reports grouped by source ID. Each mapper run has three comparisons: the dependency read, the range read and the outer box read. The checker tests the instantiated right endpoint first, then checks the old left endpoints and surplus fields. It agreed on all 828 occurrence decisions. Each fraction in Table~\ref{tab:mixed-occurrences} counts changed reports out of three occurrences per release pair. With the full row first, the seven pairs containing removals all reselected the dependency type. Only the dependency-read report changed on each pair. Thus fourteen of the twenty-one occurrence reports in that group remained reusable. The field-first catalogue preserved all twenty-one. Across both policies 810 occurrence reports were preserved. Of the 18 projections with a changed report, each changed exactly one of three occurrences. Those projections contain 54 occurrence reports. Reconstruction identifies 18 that changed and 36 that can carry forward their previous review. Table~\ref{tab:mixed-occurrences} also partitions the 18 changes by the first failed check. The required type changed in seven, and an old leaf endpoint changed at a fixed required type in eleven. Omitting the surplus check would therefore give the same 828 decisions in this cohort. On the Chalk deletion, the same omission would incorrectly accept the old report.

\input{sections/support/mixed-edits-table.tex}

\paragraph{Removing the surplus test.}
We enumerated the 133 ground types used in the cut study. For each required type \(u\), we paired every old and new left type that both successfully compared with \(u\), including equal pairs. Of the 100{,}273 ordered triples, 95{,}851 were outside the width-extension relation. Full reconstruction preserved 9{,}749 reports. We also replayed the expansion rule at every old internal node, with the same endpoint tests as the criterion. Across 2{,}357 successful old comparisons, full replay scheduled 2{,}436 internal branch tests. Corollary~\ref{cor:surplus} retained 196, reducing the branch count in 1{,}736 old comparisons. These counts sum test positions once per old comparison, before any new input is checked. Full replay and the reduced criterion agreed with reconstruction on every triple. Dropping only the surplus tests falsely accepted 196 changed reports. Among the old comparisons in this enumeration, 2{,}161 needed no surplus test and 196 needed one.

\paragraph{Independent deletion witnesses.}
A separate control has two branches that both require a surplus field. Its required type is \(\{a:\{k:Int\}, b:\{k:Int\}\}\). Its old left type adds \(x:Bool\) under \(a\) and \(y:Bool\) under \(b\). The old report has an \(Int\) equality at each of \texttt{a.k} and \texttt{b.k}. Deleting only \(x\) leaves both old paths typed \(Int\). The new report stops at whole-record equality at \texttt{a} and keeps the old leaf at \texttt{b.k}. Deleting only \(y\) keeps \texttt{a.k} and moves the other stop to \texttt{b}. Each deletion passes the endpoint tests and the sibling's surplus test. Omitting the test on the edited branch would therefore accept a changed report in each case.

%% file: sections/support/mixed-edits-table.tex
\begin{table}[!htbp]
\centering\small
\caption{Changes to individual comparisons across npm releases}\label{tab:mixed-occurrences}
\begin{tabular}{@{}lrrrrrr@{}}
\toprule
Read policy & Removal (7) & Other (131) & All (138) & Required & Leaf & Surplus\\
\midrule
Full row first & 7/21 & 11/393 & 18/414 & 7 & 11 & 0\\
One field first & 0/21 & 0/393 & 0/414 & 0 & 0 & 0\\
\bottomrule
\end{tabular}
\end{table}

%% file: sections/support/rush-reports.tex
\subsection{Comparison reports across API Extractor releases}\label{sec:rush-review}

Rush Stack keeps API Extractor's public declarations in a versioned report and names owners for changes to the review files \citep{rush-policy,rush-owners}. The configuration reader consumes two of those declarations: \texttt{IConfigApiReport} controls which API reports to write, and \texttt{IConfigDocModel} controls the documentation model. A pull request adding support for reports by release level changed the configuration reader and the tracked API report and received a recorded approval \citep{rush-variants}. The upstream approval covers the public API. We use the two configuration interfaces to track the effect of signature changes on the comparisons in our mapper.

We took every stable API Extractor release from 7.40.0 up to, but excluding, 7.50.0, in npm publication order. For each of the 47 releases we followed its published Git commit to the tracked report \citep{rush-cohort}. We read the two interfaces named by \texttt{IConfigFile.apiReport} and \texttt{IConfigFile.docModel}. Our descriptor retains each declared member name and encodes its optional marker and written type as one distinct label in a singleton record. Thus a change from an optional to a required member, or a change in its written type, changes its descriptor. Type references are opaque. The mapper reads the two descriptors at separate source occurrences. Both read types are fixed to their respective declarations in 7.40.0. The upstream report and code identify the interfaces; the two comparison reports are produced by GCP.

For each adjacent pair, we compared two predictions with fresh Java reconstruction. The source-path test derives the protected paths of each occurrence from the earlier certificate and checks where fields were added. The member-diff baseline flags a comparison whenever its interface's map of member names to written signatures changes. All 46 pairs admitted both fixed read types and satisfied width extension. In 45 pairs both descriptors stayed the same; one pair added a member. Each method agreed with Java on all 92 decisions, identifying one changed and 91 preserved comparisons (Table~\ref{tab:rush-review}). The mixed-edit check on the same certificates agreed throughout. No tracked member was deleted or replaced in this window.

\begin{table}[!htbp]
\centering\small
\caption{Affected comparisons across API Extractor releases}\label{tab:rush-review}
\begin{tabular}{@{}lrrrrrr@{}}\toprule
Window & Pairs & Decisions & Java changes & Member alerts & Path alerts & Member only\\\midrule
Original & 46 & 92 & 1 & 1 & 1 & 0\\
Continuation & 61 & 122 & 1 & 2 & 1 & 1\\
\bottomrule\end{tabular}
\end{table}

In 7.41.0 four exported enums lost their \texttt{const} modifier elsewhere in the API file. Both selected comparisons survived. In 7.44.0 the API report configuration gained \texttt{reportVariants}. Its old whole-row equality became five comparisons of the required members. The documentation-model equality survived.

\paragraph{The subsequent release history.}
We continued from 7.49.2 through every stable release in the next ten minor lines, ending at 7.59.4 before 7.60.0. We used the same mapper, the same two read types from 7.40.0 and the same member-diff rule. For each version we extracted both interfaces from the declaration file in the published npm package. The declarations agree with the tracked-report descriptors on all 47 overlapping releases. All 61 subsequent pairs satisfied width extension. Each row of Table~\ref{tab:rush-review} counts alerts per source occurrence; Java reconstructs the two reports on both sides. Every path alert matched a changed report and every changed report produced a path alert.

In 7.51.0, \texttt{releaseTagsToTrim} was added to the documentation interface. Its comparison had still stopped at equality, so the member and path checks both flagged it. In 7.52.3, \texttt{tagsToReport} was added to the API report interface \citep{rush-followup}. The earlier addition of \texttt{reportVariants} had already opened that comparison at the interface row. The new member lies outside the five required member subtrees. The member map changed, but both occurrence reports survived in fresh Java projections. Across the continuation this was the single disagreement in 122 decisions. It concerns the two reports in our mapper; Rush Stack's public-API review still covers the added member itself.

\paragraph{A controlled second addition.}
A separate control adds \texttt{unusedFreshOption}, an optional Boolean, to the 7.49.2 API report descriptor at the original read types. The member check flags the edited interface, while source prediction and fresh reconstruction preserve both reports. The control is excluded from both release windows in Table~\ref{tab:rush-review}.

%% file: sections/support/source-reuse.tex
\subsubsection{Reusing reports from an existing order pipeline}\label{sec:report-reuse}

We added a report consumer to the existing \texttt{shop.orders} Flowable example, whose two mappers are \texttt{o -> o.id} and \texttt{o -> o.sku}. The six-order program filters by amount, takes three IDs and explicitly requests two SKUs.  The example predates this integration; the report consumer and the two input revisions below were designed for this study. The integration exercises report import across fresh producer and receiver processes.

\texttt{ReuseWorkflow} runs the complete program and exports the two mapper ASTs before their separate projection. A producer JVM saves five reports for the consumed inputs. Three fresh receiver JVMs load these files, parse the current module, capture its current records and obtain new successful certificates. Import checks the mapper lexeme and report key, then renames only the session prefix; occurrence suffixes, paths, endpoints and multiplicities are untouched. The five saved reports retain their original byte sequences. Each receiver audits the imported payload against its \emph{current} export and substitution. Rejection marks the saved report as inapplicable to the current call. The original report remains associated with its producer record. Each changed input triggers a fresh projection before the receiver evaluates report reuse.

The new downstream task imports each mapper's output cell using the saved component description. It accepts only one leaf at the mapper's field path with equal \(Int\) or \(String\) endpoints. Each policy writes a separate typed-cell result and receipt. Unchecked reuse, B/M plus reuse, C-gated reuse and B plus regeneration receive the same saved reports and current calls. A full-context fingerprint compares mapper text, actual values and ordered domains. A concrete-value baseline applies value typing to the actual value at the mapper field; on a miss, regeneration is required. Python evaluates both policies on the saved same-input records.

Table~\ref{tab:report-reuse} counts a value-type hit when the concrete types at the mapper field agree. An exact-context hit also requires matching actual values and ordered domains.

\begin{table}[t]\centering\small
\caption{Reuse of saved mapper reports}\label{tab:report-reuse}
\begin{tabular}{@{}lrrrrr@{}}
\toprule
Receiver input & Reports & C & Value types & Exact context & Import failures\\
\midrule
Unchanged orders & 5 & 5 & 5 & 5 & 0\\
Add unused \texttt{note} field & 5 & 5 & 5 & 0 & 0\\
Change IDs to \texttt{ORD-102}, etc. & 5 & 2 & 2 & 0 & 3\\
\bottomrule
\end{tabular}
\end{table}

B and M accept all 15 submissions. The three stale ID reports make unchecked integer-cell import fail on string IDs; C blocks those imports and retains the SKU reports. Regeneration produces 15 correct cells. The concrete read-value type baseline achieves the same 5/5/2 reuse counts as C, including the unused-field change. The nested comparison below adds a separate baseline over all selected read-node types.

Python independently reconstructs all exports, substitutions, traces and cells and decodes the saved files and receipts. The final run has four complete application executions and 20 checked mapper executions, with no result disagreements. The pipeline executes its demand handling and accumulators through the carrier; the source theorem applies to the two exported pure mappers.

\subsubsection{Repeated reads with unchanged results}\label{sec:nested-reuse}

The nested fixture uses\par\nobreak
\texttt{r -> \{a=r.inner.k,b=r.inner.k\}} with \texttt{r=\{inner=\{aux=7,k="x",unused=true\}\}}. Four distinct roots yield six leaves. A producer process saves its complete report; two receiver processes load the same bytes, obtain fresh certificates and issue separate receipts. The second receiver changes only candidate order, preferring \(\{aux:Number,k:String\}\) to \(\{aux:Int,k:String\}\) for each intermediate field variable. Both orders are admissible. Both executions return \texttt{\{a="x",b="x"\}} with type \(\{a:String,b:String\}\).

We distinguish three observations: the concrete type of each evaluated read, the type selected for its field variable in the certificate, and the final return type. Table~\ref{tab:nested-read-types} reports all four read occurrences; two intermediate types change. The concrete type at each of the four reads is the same on both receivers. These variables are reconstructed from the source export and looked up in the recorded substitution, independently of value typing.

\input{sections/support/nested-read-types.tex}

The old per-occurrence assertions contain \(Int\preceq_K Int\) at both \texttt{/field:inner/field:aux} paths. Current decomposition instead contains \(Int\preceq_K Number\) there. All paths, source roots and counts remain unchanged; both old inequalities remain true. B, M, final-return-type equality and concrete read-value type equality therefore pass, while C rejects endorsement of the saved statement. The unchanged receiver verifies it. Direct regeneration produces a correct current report in both cases, but marks it \texttt{Regenerated}, preserving the distinction between replacement and endorsement. Both exact-context invalidation and comparison of all four selected read-node types detect the change.

The consumer attaches the review to the selected projection record. The receiver owns the current candidate profile, executes fresh projection and saves \(E,\rho,\eta,j\). Endorsement marks the old per-occurrence comparisons as applicable to that record. Regeneration with exact comparison supplies the same verdict as C. The unchanged selected types in the width controls below can still yield changed cuts.

Python reconstructs each canonical leaf and retrieves its original root's expression and location. It compares all six explanation fields, their order and multiplicity in both binary receipts and published JSON, for audit and regeneration. The three processes save six receipts. Twenty-eight well-formed negative controls alter each of occurrence, expression, location, path, lower endpoint, upper endpoint, or a repeated entry across the four output surfaces; all are rejected while claim ID, verdict and entry count stay fixed. The unmodified outputs pass. These controls validate each field and the multiplicity checks in the independent output checker.

\subsubsection{Width edits and decomposition cuts}\label{sec:frame-controls}

Eight paired parsed-source controls fix \texttt{r -> r.k}, the source export and the ordered domain within each pair. Both actual types occur in that domain; the selected field candidate is first and feasible for both calls. All 16 projections pass B. Table~\ref{tab:frame-controls} compares reuse decisions. Exact context equality rejects all eight pairs. Frame checks the old footprint, and R+compare rebuilds and compares the complete payload. Every pair preserves its selected read type and final return type. Only the first three and last pairs also preserve the concrete read value; the nested width edits extend that returned record.

\begin{table}[t]\centering\small
\caption{Reuse decisions on eight width extensions}\label{tab:frame-controls}
\begin{tabular}{@{}lrrrr@{}}
\toprule
Old comparison / edited region & Pairs & Frame & C & R+compare\\
\midrule
Equal root / add \texttt{note} & 1 & 0 & 0 & 0\\
Proper root / add \texttt{note} & 1 & 1 & 1 & 1\\
Proper root by \(Int\preceq Number\) / add \texttt{note} & 1 & 1 & 1 & 1\\
Equal nested record / add field within cut & 1 & 0 & 0 & 0\\
Proper nested record / add ignored field & 1 & 1 & 1 & 1\\
Empty required row / extend its left endpoint & 1 & 0 & 0 & 0\\
Top cut / extend its left endpoint & 1 & 0 & 0 & 0\\
Unread subtree / extend that subtree & 1 & 1 & 1 & 1\\
\bottomrule
\end{tabular}
\end{table}

In the first pair, \texttt{\{k=7\}} becomes \texttt{\{k=7,note=true\}}. The old leaf is the equal whole-record pair at the empty path; the new leaf is \(Int\preceq_K Int\) at \texttt{/field:k}. Hence unused-field extension does not universally preserve reports, even with all selected read types fixed. An existing extra field or the strict depth comparison \(Int\preceq_K Number\) keeps the old root open and allows the same addition. Nested equality, empty-row and top cuts protect their whole left endpoints.

An independent type-level checker computes the footprint only from the old pair and compares its protected subtrees in the new type. It reconstructs both Java reports separately to test the prediction, including session/root identity, paths, endpoints and truth values. A bounded universe of 133 primitive/record types contains 518 width-extension pairs and 4,422 admissible old-comparison/new-left triples. All satisfy Theorem~\ref{thm:report-frame}: 3,134 preserve reports and 1,288 change them. The same triples agree with the record-only criterion of Corollary~\ref{cor:record-cuts}. Section~\ref{sec:report-frame} proves the result for all types in the stated fragment.

\input{sections/support/frame-evaluation.tex}

%% file: sections/support/nested-read-types.tex
\begin{table}[t]
\centering\small
\caption{Concrete and selected types for nested reads}\label{tab:nested-read-types}
\begin{tabular}{@{}llll@{}}
\toprule
Read occurrence & Concrete, both & Producer selected & Changed selected\\
\midrule
1. \texttt{r.inner} & \(\rho_V\) & \(\rho_I\) & \(\rho_N\)\\
2. \texttt{r.inner.k} & \(String\) & \(String\) & \(String\)\\
3. \texttt{r.inner} & \(\rho_V\) & \(\rho_I\) & \(\rho_N\)\\
4. \texttt{r.inner.k} & \(String\) & \(String\) & \(String\)\\
\midrule
Final return & \(\{a:String,b:String\}\) & \(\{a:String,b:String\}\) & \(\{a:String,b:String\}\)\\
\bottomrule
\end{tabular}
\par\smallskip
\(\rho_I=\{aux:Int,k:String\},\quad\rho_N=\{aux:Number,k:String\}.\)\par
\(\rho_V=\{aux:Int,k:String,unused:Bool\}.\)
\end{table}

%% file: sections/support/frame-evaluation.tex
\subsubsection{Compositional edits and reporting policies}\label{sec:frame-composition}

Eight further pairs connect the source corollary to the receiving interface (Table~\ref{tab:frame-composition}). Each pair uses one parsed export and one ordered candidate vocabulary for both calls. Actual types are width-extended, while every selected read-variable binding remains fixed. Both certificates pass B. Independent reconstruction checks the original roots, substituted lower endpoints, returns, pure values, cut footprints and every published receipt entry. The Java audit accepts five saved reports and rejects three, exactly as the footprint predicts. Regeneration produces the current report in every pair. The paired runs use the production exporter, finite solver and receiving interface.

Table~\ref{tab:frame-composition} codes equality as 1 and change as 0.

\input{sections/support/frame-composition-table.tex}

The duplicate cases retain two occurrences with identical inequalities. The nested cases retain four roots, including each intermediate read. \texttt{r -> \{box=r.k,unused=r\}.box} constructs a lower endpoint containing both a selected read type and the changing parameter type; only the former is protected. In \texttt{r -> \{box=r\}.box}, the old whole-record equality instead protects the complete constructed endpoint. The three-parameter case \texttt{r -> s -> t -> \{a=r.k,b=s.k,c=t\}} delivers captures in groups $(1,1,1)$, $(1,2)$, $(2,1)$ and $(3)$, saving a fresh immutable closure at each group boundary. All eight before/after group runs have the same respective final certificates and reports; intermediate snapshots preserve every original root. Together with the seven one-parameter pairs this gives 22 checkpoint runs over 16 distinct completed contexts. For \texttt{r -> r}, the return type changes with the input while the empty report remains equal.

Always expanding nonempty record pairs preserves all eight reports in these compositional pairs. On the original eight single-read pairs it preserves six, compared with four under the current stopping policy: it additionally preserves the root and nested equality cases, but the empty-row and top witnesses still change. The expanded report loses the whole-record equality assertion unless the consumer stores it separately. Payload size can move in the opposite direction. In the 256-field equal-subtree control, the current report has one leaf and the expanded policy has 256; their serialized sizes and reuse verdicts appear in Table~\ref{tab:frame-policy}. The edit adds one field, and the recorded sizes include both endpoints of each old leaf. Both policies describe successful comparisons, with a different unit of saved assertion.

\input{sections/support/frame-policy-table.tex}

\input{sections/support/consumer-evaluation.tex}
\input{sections/support/approval-replay.tex}

\input{sections/support/frame-receiving.tex}

%% file: sections/support/frame-composition-table.tex
\begin{table}[t]
\centering\small
\caption{Report preservation under compositional edits}\label{tab:frame-composition}
\begin{tabular}{@{}lrrrr@{}}
\toprule
Pair & Roots & C/footprint & Expanded & Return equal\\
\midrule
Repeated reads / open & 2 & 1 & 1 & 1\\
Repeated reads / equality & 2 & 0 & 1 & 1\\
Nested reads / open & 4 & 1 & 1 & 1\\
Nested reads / equality & 4 & 0 & 1 & 1\\
Construct then read / open & 2 & 1 & 1 & 1\\
Construct then read / equality & 1 & 0 & 1 & 1\\
Three captures / four groupings & 2 & 1 & 1 & 0\\
Rootless identity & 0 & 1 & 1 & 0\\
\bottomrule
\end{tabular}
\end{table}

%% file: sections/support/frame-policy-table.tex
\begin{table}[t]
\centering\small
\caption{Report size and reuse under two stopping policies}\label{tab:frame-policy}
\begin{tabular}{@{}lrrr@{}}
\toprule
Policy & Old leaves & Old KiB & Reusable\\
\midrule
Stopping position & 1 & 10.56 & 0\\
Expand records & 256 & 16.75 & 1\\
\bottomrule
\end{tabular}
\end{table}

%% file: sections/support/consumer-evaluation.tex
\subsubsection{Checking a delivery approval}\label{sec:consumer-evaluation}

The consumer of Section~\ref{sec:consumer-contract} stores one approved
leaf. It replays only that path for comparison evidence, and an
independent shape interpreter reads the capture types and the frozen
source expression. We compare three policies on the same completed
certificates and captures. \emph{Selected} checks the approved leaf
and exact shape. \emph{Whole} checks the submitted report against the
report bound at preparation, compares it with a reconstruction from the
current certificate, and checks exact shape. \emph{Direct} evaluates
the body and checks the resulting value against the same closed schema.
Each policy checks B before authorizing delivery. The review policies
use a prepared approval, and direct validation needs only the schema
and the current context.

\paragraph{Two separating regressions.}
For \texttt{r -> \{box=\{k=r.k\},spare=r\}.box}, adding \texttt{note}
to the input changes the comparison for the inner \texttt{r.k}
from equality to field decomposition. The outer \texttt{/field:box}
assertion remains \(\{k:Int\}=\{k:Int\}\) and the result remains
\texttt{\{"k":7\}}. Whole-report checking rejects the changed report.
Selected checking and direct validation both deliver that result.

For \texttt{r -> \{box=r.inner,spare=true\}.box}, the selected type
for \texttt{r.inner} is \(\{k:Int\}\), even when the actual
\texttt{inner} contains \texttt{note}. On two identical widened
captures, the earlier equality-only release rule verifies the full
report and emits \texttt{\{"k":7,"note":true\}}. Preparing an approval
under the new rule fails because the exact shape has the extra field.
A separate edit starts with a valid prepared approval and adds
\texttt{note} inside \texttt{inner}. The approved leaf still matches
and the output shape fails. These runs distinguish unsound preparation
from invalidation of an existing approval. In a further control,
\texttt{r -> r.box} loses a spare input field. The comparison now stops
at root equality, the approved field leaf disappears, and the two
review policies block a value that passes direct validation.

In the unrelated-cut case, replacing the saved report with the current
canonical report under the same ID yields \texttt{REPORT\_SUBSTITUTED}
and writes no file. In the encoding control, a string containing an
unpaired UTF-16 surrogate passes B, report and shape checks.
All three policies return \texttt{ENCODING\_FAILED} and write no file.

In Table~\ref{tab:consumer-outcomes}, Leaf, Report, Shape and Encoding name the check that stopped delivery. No approval means that preparation rejected the old context.

\input{sections/support/consumer-outcomes.tex}

\paragraph{Direct assertion checking.}
Table~\ref{tab:predicate-controls} applies the explicit predicate of
Section~\ref{sec:approval-contract-choice} to the same approved equality
and frozen certificate as the leaf test. The new-ancestor-cut control
retains that predicate while losing the leaf. The unrelated-cut control
retains both while changing the report. The widened-intermediate control
retains both but fails exact shape. Thus stricter evidence preservation
and payload compliance answer separate questions. A consumer choosing
the predicate contract can accept the ancestor-cut edit after the common
B, shape and encoding checks; requiring leaf identity deliberately rules
out that reuse.

An independent reconstruction checks the predicate, prefix guards and report membership on all eight consumer cases, including the four in Table~\ref{tab:predicate-controls}. It also checks the trace-replay factorization on the bounded type universe used for the footprint checks.

\input{sections/support/predicate-controls.tex}

\paragraph{Same-input costs.}
Table~\ref{tab:consumer-cost} measures the in-memory path from a current
certificate to a decision and, on delivery, serialized JSON. The
functional runs also write and read back the delivered JSON files.
Timing includes original-bound checking, policy checks, evaluation
when needed, and encoding. Parsing, projection, approval preparation,
claim decoding and file writes precede or follow the measured path.
The unsafe preparation and encoding failure are functional controls.
The other six cases each contribute seven shuffled blocks of 200
invocations per operation in each of three sequential JVM forks, after
300 warmups per operation. Python reconstructs both reports and exact
shapes and checks the decisions and emitted payloads.

Tables~\ref{tab:consumer-cost} and~\suppnumber{tab:consumer-stages} give the median of three fork medians in microseconds for complete delivery and individual operations. Tables~\suppnumber{tab:consumer-cost-ranges} and~\suppnumber{tab:consumer-stage-ranges} give the ranges of the three fork medians for complete delivery and for its individual operations, respectively, in the same units.

\input{sections/support/consumer-cost.tex}
\input{sections/support/consumer-cost-reading.tex}

The three Selected fork medians in the value-only case span
2.26--14.03 microseconds. Section~\suppnumber{app:consumer-cost} reports these
ranges and measures the individual operations.

%% file: sections/support/consumer-outcomes.tex
\begin{table}[t]\centering\small
\caption{Delivery outcomes by approval policy}\label{tab:consumer-outcomes}
\begin{tabular}{@{}llll@{}}\toprule
Edit or control & Selected & Whole & Direct\\\midrule
Value only & Deliver & Deliver & Deliver\\
Extra output field & Leaf & Report & Shape\\
Unrelated cut & Deliver & Report & Deliver\\
Widened intermediate & Shape & Report & Shape\\
Unsafe old context & No approval & No approval & Shape\\
New ancestor cut & Leaf & Report & Deliver\\
32 sibling reads & Deliver & Deliver & Deliver\\
Invalid wire string & Encoding & Encoding & Encoding\\
\bottomrule\end{tabular}\end{table}

%% file: sections/support/predicate-controls.tex
\begin{table}[t]\centering\small
\caption{Evidence and output shape in four delivery controls}\label{tab:predicate-controls}
\begin{tabular}{@{}lllll@{}}\toprule
Control & Predicate & Leaf & Whole & Exact shape\\\midrule
Value only & Yes & Yes & Yes & Yes\\
Unrelated cut & Yes & Yes & No & Yes\\
New ancestor cut & Yes & No & No & Yes\\
Widened intermediate & Yes & Yes & No & No\\
\bottomrule\end{tabular}\end{table}

%% file: sections/support/consumer-cost.tex
\begin{table}[t]\centering\small
\caption{Delivery time by approval policy}\label{tab:consumer-cost}
\begin{tabular}{@{}lrrr@{}}\toprule
Edit or control & Selected & Whole & Direct\\\midrule
Value only & 10.33 & 14.11 & 7.26\\
Extra output field & 2.30 & 2.91 & 1.83\\
Unrelated cut & 3.57 & 3.83 & 2.78\\
Widened intermediate & 14.75 & 17.53 & 10.00\\
New ancestor cut & 1.48 & 1.89 & 1.17\\
32 sibling reads & 32.85 & 53.24 & 24.93\\
\bottomrule\end{tabular}\end{table}

%% file: sections/support/consumer-cost-reading.tex
The selected policy delivers 3 payloads and the whole-report policy delivers 2. Direct validation delivers 4. With 32 sibling reads, all three deliver. Their medians are 32.85, 53.24, and 24.93 microseconds, respectively. Direct validation has the lowest median in 6 of the 6 timed cases. The unrelated-cut row takes different branches: the whole-report policy rejects, and the other two evaluate and encode a payload. That row measures the costs of the actual decisions under each policy.

%% file: sections/support/approval-replay.tex
\subsubsection{Restoring approvals and reselecting bindings}\label{sec:approval-replay}

The mapper \texttt{r -> \{box=r.order,spare=r.audit.amount\}.box}
returns an order and reads its audit amount in a sibling field.
A producer uses the six \texttt{shop.orders} records to prepare
\texttt{box} approvals under the order schema and root-equality
controls for \texttt{r -> r.box}. A second JVM reparses, projects,
restores and delivers. A third runs the width survey.

\input{sections/support/approval-replay-results.tex}

The survey adds one Boolean field at three positions in each
wrapped order and freshly projects three mappers under two candidate
orders. Each pair shares its ordered domain. The six orders share
their type shapes, so each order of candidates covers nine mapper
and edit combinations repeated across six values.

\input{sections/support/approval-survey-results.tex}

\begin{figure}[tbp]
\centering
\includegraphics[width=.96\linewidth]{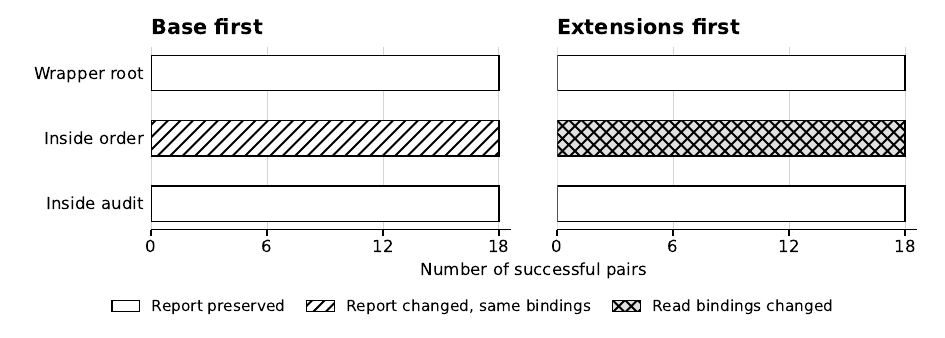}
\caption{Report outcomes by edit location and candidate order}
\Description{In both panels, all 18 wrapper-root edits and all 18 edits inside audit preserve the report. All 18 edits inside order change the report in both panels. The left panel retains the selected read bindings, while the right panel changes them.}
\label{fig:approval-width-edit-outcomes}
\end{figure}

Figure~\ref{fig:approval-width-edit-outcomes} groups the same 108 pairs by edit location. Each bar covers 18 successful pairs: three mappers applied to six order values of the same type shape. Table~\ref{tab:approval-width-survey} partitions the successful pairs into three outcomes within each candidate order. For edits inside \texttt{order}, base-first projection keeps the read bindings and changes the report, while extensions-first projection selects different read bindings. The two candidate orders have the same total number of preserved reports across these fixtures. Their changed reports arise in different ways.

An additional control starts with an existing extra field in the order
record. Adding a second field makes the earlier extended candidate
feasible. The solver selects it and the report changes. Holding the
old binding fixed still passes B and preserves the report. These
controls distinguish preservation from candidate selection. Python
independently reconstructs the exports, projections and decisions.
Over 133 bounded ground types it also checks 189,217 pairs of
successful comparisons with width-related lower types and independently
chosen upper types. Exactly 3,134 reports agree, each satisfying both
the upper-type and the old-cut tests.

\paragraph{Evolving the capture catalogue.}
A producer saves the original report once. A fresh receiver JVM
processes the six orders under a sequence of catalogues. The original
audit record contains an amount and an unused flag. The next
catalogue adds the capture shape with a second flag, and the following
catalogue adds a shape with a third. The receiver continues to use the
initial approval on both steps. The unchanged export and closed
output schema identify the same order delivery. The order values are
from the existing example; the catalogue transitions and consumer
are experimental fixtures.

\input{sections/support/approval-migration-tables.tex}

Table~\ref{tab:approval-migration} separates admission, projection,
report identity, restoration and delivery. Each case uses six orders apart from the single 32-read request. Restored counts successful restorations under the append policy. Six unlisted captures stop
before projection. The other 67 requests use the same fresh
certificate under all three policies. Exact identity delivers 6
payloads. The append policy delivers 19, including the two catalogue
additions and the 32-read case. The current catalogue policy delivers
37, including six each after capture reordering, removal of an unused
capture candidate and read-domain addition. These 18 reports remain canonical and the
payloads match the corresponding unchanged controls byte for byte.
Both migration policies reject source or schema changes, changed
reports and same-ID replacements. The cut and reconstruction routes
agree on all decisions and payloads within each migration policy. The
difference between 6, 19 and 37 is attributable to which catalogue
changes the policies admit under this shared check. It measures the
effect of each policy on these designed requests.

\input{sections/support/approval-reselection-results.tex}

A build report whose contract is to publish the current solver's
selected type of \texttt{r.audit} must reproject after the capture
changes. In these
six requests the selected row grows from \(\{\mathsf{amount:Int}\}\)
to one that also includes \texttt{existing} and \texttt{note}.
The old report cannot explain that new selection. A wire consumer that accepts any valid witness can retain the old read assignment. B establishes typing for that assignment on the current captures; the report check establishes whether it still produces the approved evidence.

A two-parameter control compares saved approvals with successive
lists \([a]\), \([a,b]\), and \([a,c]\) for one parameter. All three
contexts pass B and give the same report. The approval saved at
\([a]\) delivers under both later catalogues. An approval saved at
\([a,b]\) fails restoration under \([a,c]\) because its saved entry
\(b\) is missing.

Two further controls check withdrawal of candidates used by the
old certificate. Withdrawing the type of the actual capture blocks
admission before projection. Withdrawing a selected read type makes
the held assignment invalid in the current catalogue. Fresh projection
succeeds with a different binding and the saved report fails review.
Direct schema checking still accepts the resulting value.

Table~\ref{tab:approval-migration-cost} compares the cut route with
complete reconstruction from an ordinary approval file containing
the bound report and definition, without the old assignment.
Both routes include file reads, restoration, claim decoding, B,
report checking and, on delivery, shape checking, evaluation and
encoding. Projection and payload file writes are outside the timer.
Three JVM forks run 150 warmups and seven blocks of 40 requests per
route and case with shuffled case and route order. Ordinary
reconstruction has lower median receipt times and smaller approval
files in all four measured cases. These timings cover the append
policy after projection. Times are microseconds, summarized by the median of the three JVM fork medians. A pipeline that publishes a freshly selected certificate also incurs the projection time measured in Table~\ref{tab:approval-cycle}.

\paragraph{From an initial review to a fresh revision.}
Table~\ref{tab:approval-cycle} measures one preparation followed by one request with fresh projection. Each trial starts with the first order value in memory. Preparation parses and exports the old mapper, binds the old capture, projects a certificate, constructs its canonical claim and writes the claim and the route-specific approval to local files. The receiving phase parses and exports the mapper anew, admits the revised capture, projects a fresh certificate, reads both files and runs the corresponding receiver. On success it writes the encoded payload. The framed route checks eligibility and reconstructs the report on fallback. The extra removal case deletes \texttt{existing} from the audit row. The old comparison expanded through \texttt{amount}; the new row equals the required row. The framed route falls back and both routes withhold delivery.

\input{sections/support/cycle-table.tex}

We used the same first order, mapper and catalogue for each pair of runs on an arm64 macOS host with JDK~21. Three fresh JVMs each ran 12 warmup cycles per route and case and seven paired blocks of four cycles per route. We shuffled case order and randomized which route ran first in each block. Both routes used ordinary buffered file I/O in the same JVM, with no network transfer, disk synchronization or human review inside the timer. We checked every timed decision and successful payload against direct reconstruction. The framed route stored more bytes in every case. Bytes include the approval and the submitted claim. R denotes reconstruction and C the cut route. We report milliseconds as medians of three fork medians, with the total measured over each complete cycle. For the 32-read case both totals exceeded 430~ms. Preparation and receipt each took over 200~ms per route when each phase included a fresh projection. The first append had a lower median total for Cuts (1.319 versus 1.360~ms), with fork medians spanning 0.781--2.234~ms for Cuts and 0.831--1.990~ms for Rebuild. On removal, fallback took 0.736~ms for the full cycle versus 0.683~ms for direct reconstruction. The ordering of the median total varies with the workload. The prospective classification of insertion sites in Corollary~\ref{cor:source-footprint} holds for either choice of receiving algorithm.

%% file: sections/support/approval-replay-results.tex
The receiver restores 24 of 36 requests. An added field in the audit input changes 6 full reports. All 6 orders still carry the saved \texttt{box} approval and are delivered under single-assertion checking. Whole-report checking withholds those orders pending new approval. The receiver delivers all 6 root-equality controls and withholds every extra-output-field payload under all three policies. Changes to the mapper or candidate order require new approval.

%% file: sections/support/approval-survey-results.tex
\begin{table}[t]\centering\small
\caption{Projection outcomes by candidate order}\label{tab:approval-width-survey}
\begin{tabular}{@{}lrrrrr@{}}\toprule
Order & Both succeed & Same report & Same bindings only & Changed bindings & Same return\\\midrule
Base first & 54 & 36 & 18 & 0 & 54\\
Extensions first & 54 & 36 & 0 & 18 & 48\\
\bottomrule\end{tabular}\end{table}

Table~\ref{tab:approval-width-survey} counts the successful pairs. The fixed-binding control has 6 pairs, one per order.

%% file: sections/support/approval-migration-tables.tex
\begin{table}[t]\centering\small
\caption{Approval migration under three catalogue policies}\label{tab:approval-migration}
\begin{tabular}{@{}lrrrrrrr@{}}\toprule
Request & Admitted & Same reads & Same report & Restored & Exact & Append & Current\\\midrule
Unchanged & 6 & 6 & 6 & 6 & 6 & 6 & 6\\
Unlisted capture & 0 & 0 & 0 & 0 & 0 & 0 & 0\\
First append & 6 & 6 & 6 & 6 & 0 & 6 & 6\\
Second append & 6 & 6 & 6 & 6 & 0 & 6 & 6\\
Protected subtree & 6 & 6 & 0 & 6 & 0 & 0 & 0\\
Read reselection & 6 & 0 & 0 & 6 & 0 & 0 & 0\\
Capture reorder & 6 & 6 & 6 & 0 & 0 & 0 & 6\\
Unused capture candidate removed & 6 & 6 & 6 & 0 & 0 & 0 & 6\\
Read-domain append & 6 & 6 & 6 & 0 & 0 & 0 & 6\\
Same-ID replacement & 6 & 6 & 0 & 6 & 0 & 0 & 0\\
Source edit & 6 & 6 & 0 & 0 & 0 & 0 & 0\\
Schema edit & 6 & 6 & 6 & 0 & 0 & 0 & 0\\
32 sibling reads & 1 & 1 & 1 & 1 & 0 & 1 & 1\\
\midrule
Total delivered & & & & & 6 & 19 & 37\\
\bottomrule\end{tabular}\end{table}

\begin{table}[t]\centering\small
\caption{Storage and receiving costs under the append policy}\label{tab:approval-migration-cost}
\begin{tabular}{@{}lrrrr@{}}\toprule
Request & Ordinary bytes & Framed bytes & Rebuild & Cuts\\\midrule
First append & 2,493 & 2,987 & 120.01 & 133.45\\
Protected subtree & 2,502 & 2,968 & 77.38 & 90.65\\
Read reselection & 2,873 & 3,367 & 101.08 & 102.14\\
32 sibling reads & 32,060 & 34,780 & 473.38 & 524.07\\
\bottomrule\end{tabular}\end{table}

%% file: sections/support/approval-reselection-results.tex
With the same current captures and approval, retaining the old read assignment gives a B-valid certificate, the original report and the same exact output shape in all 6 cases. Both checks deliver all 6 payloads with this certificate. Fresh projection delivers 0 of 6 because its selected report changes.

%% file: sections/support/cycle-table.tex
\begin{table}[tbp]\centering\small
\caption{Cost of a complete approval cycle}\label{tab:approval-cycle}
\begin{tabular}{@{}lrrrrrrrr@{}}\toprule
& \multicolumn{2}{c}{Stored bytes} & \multicolumn{2}{c}{Prepare} & \multicolumn{2}{c}{Receive} & \multicolumn{2}{c}{Total}\\
Case & R & C & R & C & R & C & R & C\\\midrule
First append & 3,197 & 3,691 & 0.677 & 0.633 & 0.691 & 0.661 & 1.360 & 1.319\\
Protected subtree & 3,244 & 3,710 & 0.375 & 0.384 & 0.298 & 0.325 & 0.683 & 0.716\\
Read reselection & 3,577 & 4,071 & 0.474 & 0.458 & 0.420 & 0.425 & 0.900 & 0.895\\
32 sibling reads & 39,877 & 42,597 & 209.163 & 210.567 & 222.989 & 223.299 & 432.152 & 434.046\\
Field removal & 3,197 & 3,691 & 0.381 & 0.411 & 0.298 & 0.321 & 0.683 & 0.736\\
\bottomrule\end{tabular}\end{table}

%% file: sections/support/frame-receiving.tex
\subsection{Whole-report file receiving}\label{sec:frame-receiver}

We implement Section~\ref{sec:frame-receiver-contract} through the existing claim-file adapter. Its preparation step reads and verifies an old claim and retains its record-valued cuts. Receiving checks file separation, reads the input, tests the retained context, checks the current certificate, and either applies the footprint or reconstructs and compares. Both paths encode and write a separate receipt. Seven cases cover value-only change, an added field, a removed field, a changed selected binding, an invalid returned type, an altered claim and 32 repeated reads. Every resulting receipt is byte-identical to direct reconstruction and comparison. Table~\ref{tab:frame-receiving} gives the median of three JVM fork medians in microseconds. Table~\suppnumber{tab:frame-receiving-ranges} gives the corresponding ranges.

The original source-specific release example uses \texttt{r -> \{box=r,spare=true\}.box} with selected result \(\{k:Int\}\). The consumer has a closed wire contract permitting exactly one integer field \texttt{k}. Its reviewed assertion is the equality of the actual and required records at \texttt{/field:box}, associated with the outer source read. It releases a serialized payload only if the current verified receipt still contains that assertion. Changing 7 to 9 preserves it and writes \texttt{\{"k":9\}}. Adding \texttt{note} changes the actual returned record and the assertion, so the receiving task withholds the output file. The static return type and fully expanded field reports remain equal across the addition. In separate output files, using either equality as the approval gate delivers both payloads, including the one that violates the schema. Direct exact-shape checking and the retained whole-record assertion both withhold that payload. For this expression the compared left field is the exact type of the captured value. Section~\ref{sec:consumer-evaluation} tests the more general consumer contract.

For the other five cases, field removal and changed read binding fail the footprint guard and invoke reconstruction. An altered claim also invokes reconstruction. Invalid return evidence produces InvalidCertificate. The 32-read case has no record-valued cuts, so successful premise validation decides report preservation. Python reconstructs the original bounds, ground assignments, both report policies, full binary receipts and serialized output, and checks every route.

\input{sections/support/frame-receiving-table.tex}

\paragraph{Measured receiving cost.}
Three sequential JVM forks each run 100 warmups per operation followed by seven shuffled blocks of 20 invocations per case, giving 147 observations. Cases and operation order are shuffled with fixed seeds. Preparation includes reading and validating the retained claim and building the footprint. First includes preparation and a receiving invocation in one measured operation. Repeated receiving starts with that prepared state and includes original-bound checking, applicability tests, file input and receipt output. The baseline reads the same input and directly reconstructs and compares, without a width-extension guard. Parsing, capture, certificate search, network transport and durability synchronization precede these timed operations or belong to another task. Payload delivery is checked separately after the receipt decision.

\input{sections/support/frame-receiving-reading.tex}

Section~\ref{sec:frame-cost} summarizes the ground comparison experiment. Section~\suppnumber{app:frame-cost} retains measurements for all twelve cases. In that earlier full-cut prototype, a fresh footprint query with validation should be compared against direct reconstruction R. At width 256 only the many-field row has a lower median, 35.18 versus 53.03 microseconds. Sparse reads take 18.94 versus 0.25, and the intact equality-cut case takes 30.16 versus 19.24. These results motivate testing the reduced criterion in the receiving adapter, where input and receipt costs are included.

\paragraph{Fixed-catalogue restart control.}
Table~\ref{tab:bound-frame} uses one extended record format for both
routes. The producer saves the old assignment and exact report.
The receiver checks nine requests, including narrowing, substitution
under the same ID, a bad return certificate and an encoding failure.
The two routes agree on every decision and payload. A predeclared-domain
control admits one audit extension, whose report changes, and refuses
two unlisted shapes. The measured boundary includes reading and
restoring the approval and encoding a delivered value, with 189 paired
blocks in three JVMs. Section~\ref{sec:approval-replay} uses a separate
ordinary-record baseline for catalogue changes.
Both routes return the decision shown in Table~\ref{tab:bound-frame}. The two time columns give the median of three JVM fork medians in microseconds.

\input{sections/support/approval-frame-table.tex}

%% file: sections/support/frame-receiving-table.tex
\begin{table}[t]\centering\small
\caption{File receiving time by request type}\label{tab:frame-receiving}
\begin{tabular}{@{}llrrrrr@{}}\toprule
Case & Route & Record cuts & Prep & First & Receive & Rebuild\\\midrule
wire same shape & Frame & 1 & 36.71 & 134.88 & 96.51 & 100.85\\
wire width add & Frame & 1 & 33.03 & 127.21 & 87.90 & 93.89\\
remove field & Rebuild & 0 & 38.90 & 150.50 & 106.57 & 92.13\\
changed binding & Rebuild & 1 & 22.95 & 103.62 & 84.52 & 77.34\\
invalid return & Invalid & 1 & 25.81 & 105.57 & 85.43 & 81.12\\
changed claim & Rebuild & 1 & 23.61 & 105.67 & 76.33 & 77.89\\
32 reads & Frame & 0 & 100.51 & 260.47 & 164.68 & 172.88\\
\bottomrule\end{tabular}\end{table}

%% file: sections/support/frame-receiving-reading.tex
For 32 repeated reads, receiving takes 164.68 microseconds and direct reconstruction takes 172.88. Their fork-median ranges are 157.61--167.39 and 160.06--176.83. Preparation takes 100.51, and the directly timed first request including preparation takes 260.47. Directly timed first requests have a higher median than reconstruction in 7 of the 7 cases. The small control cases and fallbacks expose the cost of checking applicability. Fork ranges overlap in several cases, so these observations support workload-specific cost estimates. Section~\suppnumber{app:frame-receiving-cost} gives every fork range.

%% file: sections/support/approval-frame-table.tex
\begin{table}[t]\centering\small
\caption{Approval decisions and timing after restart}\label{tab:bound-frame}
\begin{tabular}{@{}lllrr@{}}\toprule
Request & Decision & Criterion stops at & Criterion & Rebuild\\\midrule
Value only & Delivered & Cuts & 114.07 & 115.08\\
Outside old cuts & Delivered & Cuts & 98.60 & 88.39\\
Protected subtree & Report changed & Cuts & 72.75 & 70.64\\
Read reselection & Report changed & Binding change & 71.72 & 77.67\\
Narrowing & Report changed & Rebuild & 81.04 & 84.91\\
Same-ID replacement & Report substituted & Original report & 58.25 & 57.04\\
Invalid certificate & Invalid certificate & B check & 61.84 & 57.81\\
Encoding failure & Encoding failed & Cuts & 90.65 & 87.85\\
32 sibling reads & Delivered & Cuts & 514.79 & 489.97\\
\bottomrule\end{tabular}\end{table}

%% file: sections/09-related.tex
\section{Related Work}\label{related-work}

Report identity connects certificate checking with the provenance of individual type comparisons. Section~\ref{sec:preservation-related} gives the technical comparison with trace adaptation, incremental checking and slicing. Here we consider the checker interface and the information carried by a source occurrence.

\subsection{Certificate Checking and Translation Validation}\label{sec:certificate-related}

Proof-Carrying Code separates evidence production from checking against a consumer's safety policy \citep{necula-pcc-1997}. Translation validation checks individual transformations for semantic preservation \citep{necula-tv-2000}. Here B rechecks the original inequalities for value coverage and C checks the submitted source comparisons against the selected substitution. The receiver trusts the frozen export, ground relation, capture typing and checker. This lets it validate a certificate independently of the search and residual producer.

\paragraph{Source origins in type inference}
SuperF discusses propagating inferred types' source provenance and tracking flows involving rigid variables and skolems \citep[Section~2]{parreaux-superf-2024}. Skalpel labels generated constraints and propagates dependencies through solving, including fresh renaming and substitution for Standard ML \citep[Sections~5--6]{rahli-skalpel-2017}.

Canonical regeneration plus exact comparison implements C. Theorem~\ref{thm:residual-origins} connects the report to capture and copying; Theorem~\ref{thm:report-frame} characterizes its equality after a width edit. Stable IDs and equal read bindings still permit changes to stopping paths and endpoints.

\subsection{Structural Types and Report Identity}\label{structural-subtyping-for-records-objects-and-duck-typed-languages}

Width and depth subtyping explain why the original judgment can stay valid after a field addition \citep{pierce-records,cardelli-bounded}. Row polymorphism abstracts over unused fields \citep{wand-1987-objects,remy-1989-records}. Here an equal pair of record types yields one leaf whose endpoints name the complete records. A policy that expands all nonempty records instead can preserve its field comparisons while relinquishing this whole-record assertion. The stopping rule thus determines whether the report contains the whole-record assertion or its field comparisons.

\subsection{Constraint Flow and Diagnostic Evidence}

Type error slicing labels constraints with their source locations and computes a minimal unsatisfiable subset that explains failed inference \citep{haack-wells-2004}. Skalpel extends this approach to Standard ML \citep{rahli-skalpel-2017}. GCP labels successful comparisons by source occurrence. The observation preserved across revisions includes the decomposition path and both endpoints at every leaf. Theorem~\ref{thm:report-frame} characterizes the width extensions that preserve that observation.

Getting into the Flow retains constraint-flow information and evaluates explanations with users \citep{bhanuka-flow-2023}. Its qualitative study reports demand for flow-based messages, alongside inconclusive quantitative results for error localization and understanding. GCP links the original obligation, residual specialization and published explanation through the identity of the source occurrence. A backward slice for the returned value explains the two field reads in Section~\ref{sec:preservation-related} under the observed run. Corollary~\ref{cor:occurrence-impact} instead ranges over future width extensions of that capture and identifies the exact subset of successful comparisons whose evidence will differ for each batch. A slice of the comparator must account for its equality stops and for the certificate supplied to it. Corollary~\ref{cor:surplus} identifies a finer distinction within those stops. Once the required type and old leaf endpoints agree, a record with equal required children exposes only the continued existence of some extra field. Its name and type can both change. If a required child still expands, its unequal endpoints already force the parent to expand. This is why the local test needs a surplus-field check only where all required children were equal.

%% file: sections/10-discussion.tex
\section{Discussion and Conclusion}\label{discussion-and-conclusion}

\subsection{Applicability}\label{limitations}

The source contract trusts the export, ground relation, capture typing and checker. The report-preservation theorem applies to a fixed source export, the stated ground record fragment and width extensions of captures. For arbitrary edits between two valid ground instances of the same export, Corollary~\ref{cor:surplus} instead consumes the current comparison endpoints. The finite kernel accepts a broader type syntax for individual certificate checks. Table~\ref{tab:claim-correspondence} states the guarantee boundary of each component. Section~\ref{sec:route-evaluation} and Supplementary material S1 document implementation checks, receiving costs and the scope of the evaluation.

A whole-report approval is appropriate when another party can retrieve and cite the published derivation by its review identifier. For the inventory owner who needs only the returned range, the returned-field approval has narrower scope. The Rush workflow supplies a real approval of public declarations; our mapper supplies the source comparisons. In the original release window the member and path checks made identical decisions. In the subsequent window one published member addition left both comparisons unchanged. These observations establish a difference between the two decisions on released software. The frequency and cost of human re-review would require observations of maintainers using such a contract.

Before projection, the source export, old capture types and read bindings determine the input paths relevant to a proposed insertion, provided the bindings remain fixed. After projection, a receiver can compare the new report with its saved copy or check the verified edits against those paths. In all four post-projection tasks in Table~\ref{tab:approval-migration-cost}, reconstruction used a smaller approval file and had a lower median receiving time. At width 256, direct reconstruction had a lower median than the guarded footprint query in five of the six cases in Table~\ref{tab:frame-cost}. Table~\ref{tab:approval-cycle} measures the complete cycle, including fresh projection and fallback. Tables~\ref{tab:impact-stream} and~\ref{tab:prefix-online} give the costs of maintaining the comparison reference over successive releases.

\paragraph{A release workflow with independently checked evidence.}
The Kubernetes Security Profiles Operator illustrates a release contract that binds evidence to an identified artifact. Its image promoter copies images by digest and signs a summary of the staging attestations against the project's provenance policy \citep{spo-release-verification}. Images through v1.1.0 were promoted and signed, but their builder fell outside that policy. A cluster configured to require a passing summary rejects those images. Section~\ref{sec:approval-workflow} studies a different evidence object, the ordered comparisons for a type assignment, with a receiver enforcing the selected approval contract.

\subsection{Dependency Structure of the Proofs}\label{dependency-structure-of-the-proofs}\label{app:proofs}\label{appendix-a-proofs}

Theorem~\ref{thm:residual-origins} connects the exported constraints to evaluation by substitution composition and source induction. Theorem~\ref{thm:report-frame} recovers read choices from report leaves and proves stability of the old expanding branches under width extension (Section~\ref{sec:report-frame}). The proof of Corollary~\ref{cor:source-footprint} then follows the exported lower expressions back to captures and uses single-field insertions to prove irredundancy and uniqueness under its fresh-label premise. Applying the pullback before global prefix reduction gives the per-occurrence equivalence in Corollary~\ref{cor:occurrence-impact}. For arbitrary successful endpoint edits, Corollary~\ref{cor:surplus} uses the recovered right type, cut equality and the presence of a surplus field at the remaining old branches. The delivery proof combines the applicable report test with independent checks of approval binding and output shape (Section~\ref{proof:consumer-approval}).

\subsection{Future Work}\label{future-work}

Extending the audit through body applications, mutable storage and receiver methods requires export and closure rules. Carrying approvals across source changes requires a correspondence and invalidation policy. Such a correspondence would let a reviewer follow a comparison across edits to the mapper itself.

\subsection{Conclusion}\label{conclusion}

Two source expressions can give the same old report and respond differently to the same edit. On input \(\{v=0\}\) with read type \(\{v:Int\}\), both \texttt{r -> \{k=r\}.k} and \texttt{r -> \{k=\{v=0\}\}.k} report the same whole-record equality. Adding \texttt{note} to the input changes the first report because its \texttt{k} contains the input. In the second expression \texttt{k} is constructed from a constant, and the report survives. The stopping position identifies the compared subtree. The source export supplies its connection to the input.

For a fixed export and read bindings, the reduced input paths of Corollary~\ref{cor:source-footprint} decide preservation of the complete report under field additions. Retaining the paths for each occurrence identifies the comparisons that change (Corollary~\ref{cor:occurrence-impact}). In the nested mapper, insertions at \(r.k\) and \(r.k.v\) fail the same global test, but only the deeper insertion changes both comparisons. For revisions that include deletions or new read choices, Corollary~\ref{cor:surplus} checks each occurrence against the new valid certificate. The source evidence thus supports both planning an insertion at fixed read bindings and identifying which entries of a completed revision still carry the same evidence.